\documentclass[submission,copyright,creativecommons]{eptcs}
\providecommand{\event}{PROLE 2015} 
\usepackage{etex}
\usepackage{tikz}
\usetikzlibrary{calc} 						
\usetikzlibrary{shapes,arrows}				
\usetikzlibrary{shapes.multipart}			
\usetikzlibrary{positioning}
\usepackage[all,cmtip]{xy}					
\usepackage{tikz-qtree,tikz-qtree-compat}	
\usepackage{breakurl}

\usepackage{url}
\usepackage{amssymb}
\usepackage{amsmath}
\usepackage{graphics}

 \newtheorem{definition}{Definition}
 \newtheorem{theorem}{Theorem}
   
 \newtheorem{example}{Example}  
 \newtheorem{remark}{Remark}

 \newtheorem{corollary}{Corollary}

\long\def\comment#1{}

\newcommand{\naturals}{\mathbb{N}}

\newcommand{\integers}{\mathbb{Z}}

\newcommand{\rationals}{\mathbb{Q}}

\newcommand{\reals}{\mathbb{R}}

\newcommand{\SAlgebra}{\cA}
\newcommand{\SStructure}{\cA}
\newcommand{\SModel}{\cA}
\newcommand{\SemDomain}{\cA}
\newcommand{\StructDomain}{\cA}
\newcommand{\ModelDomain}{\cA}

\newcommand{\MatrixConvexDomain}{\mathsf{C}}
\newcommand{\VectorConvexDomain}{\vec{b}}

\renewcommand{\emptyset}{\varnothing}
\renewcommand{\phi}{\varphi}

\newcommand{\pr}[1]{\mathtt{\tt #1}}

\def\defemb#1#2{\expandafter\def\csname #1\endcsname
							  {\relax\ifmmode #2\else\hbox{$#2$}\fi}}
\defemb{cA}{{\cal A}}
\defemb{cB}{{\cal B}}
\defemb{cC}{{\cal C}}
\defemb{cD}{{\cal D}}
\defemb{cE}{{\cal E}}
\defemb{cF}{{\cal F}}
\defemb{cG}{{\cal G}}
\defemb{cH}{{\cal H}}
\defemb{cI}{{\cal I}}
\defemb{cJ}{{\cal J}}
\defemb{cK}{{\cal K}}
\defemb{cL}{{\cal L}}
\defemb{cM}{{\cal M}}
\defemb{cN}{{\cal N}}
\defemb{cO}{{\cal O}}
\defemb{cP}{{\cal P}}
\defemb{cQ}{{\cal Q}}
\defemb{cR}{{\cal R}}
\defemb{cS}{{\cal S}}
\defemb{cT}{{\cal T}}
\defemb{cU}{{\cal U}}
\defemb{cV}{{\cal V}}
\defemb{cX}{{\cal X}}
\defemb{cY}{{\cal Y}}
\defemb{cZ}{{\cal Z}}

\defemb{mP}{{\sf P}}

\makeatletter
\newenvironment{prog}{\vspace{0.7ex}\par
\setlength{\parindent}{0.7cm}
\obeylines\@vobeyspaces\tt}{\vspace{0.7ex}\noindent
}
\makeatother
\newcommand{\startprog}{\begin{prog}}
\newcommand{\stopprog}{\end{prog}\noindent}

\makeatletter
\newenvironment{smallprog}{\vspace{0.7ex}\par
\setlength{\parindent}{0.7cm}
\obeylines\@vobeyspaces\tt\small}{\vspace{0.7ex}\noindent
}
\makeatother
\newcommand{\fstartprog}{\begin{smallprog}}
\newcommand{\fstopprog}{\end{smallprog}\noindent}

\makeatletter
\newenvironment{nismallprog}{\vspace{0.7ex}\par
\setlength{\parindent}{0.0cm}
\obeylines\@vobeyspaces\tt\small}{\vspace{0.7ex}\noindent
}
\makeatother
\newcommand{\fnistartprog}{\begin{nismallprog}}
\newcommand{\fnistopprog}{\end{nismallprog}\noindent}

\newcommand{\fpr}[1]{\mathtt{#1}}

\newcommand{\Maude}{{\sf Maude}}

\newcommand{\muterm}{\mbox{\sc mu-term\/}}

\newcommand{\multisolver}{\mbox{\sc MultiSolver\/}}

\newcommand{\SSymbols}{{\Sigma}}
\newcommand{\SPredicates}{{\Pi}}
\newcommand{\STermsOn}[2]{{\cT_{#1}(#2)}}
\newcommand{\STerms}{{\STermsOn{\SSymbols}{\Variables}}}
\newcommand{\GSTermsOn}[1]{{\cT_{#1}}}
\newcommand{\GSTerms}{{\GSTermsOn{\SSymbols}}}

\newcommand{\Symbols}{{\cF}}

\newcommand{\Variables}{{\cX}}

\newcommand{\Var}{{\cal V}ar}

\newcommand{\ul}[1]{\underline{#1}}

\newcommand{\toStarPosSub}[2]{{\;\mbox{$\stackrel{#1}{\longrightarrow}\hspace{.1cm}\hspace{-.2cm}^*_{#2}\,$}}}
\newcommand{\toPlusPosSub}[2]{{\;\mbox{$\stackrel{#1}{\longrightarrow}\hspace{.1cm}\hspace{-.2cm}^+_{#2}\,$}}}

\newcommand{\activationlazyrew}[1]{\stackrel{\sf A}{\to}}
\newcommand{\activationlazyrewp}[1]{\toPlusPosSub{\sf A}{}}
\newcommand{\activationlazyrews}[1]{{\toStarPosSub{\sf A}{}}}

\newcommand{\bigfrac}[2]{
\begin{array}[b]{c}
\displaystyle #1\\\hline\displaystyle #2
\end{array}}
\newcommand{\bigfracn}[3]{
\begin{array}[b]{c}
\displaystyle #1 \\\hline\displaystyle #2 
\end{array}
\hbox to 0pt{\raisebox{0.7em}{{\tiny (#3)}}}
}

\newcommand{\IFPROVED}{\mathrm{\;\; \Uparrow\;\;}}
\newcommand{\proofJump}[2]{#2\IFPROVED #1}

\newcommand{\exSymbType}[1]{\mathsf{#1}}

\newcommand{\Ffalse}{\exSymbType{false}}

\newcommand{\Ftrue}{\exSymbType{true}}

\tikzstyle{decision} = [diamond, draw, fill=yellow!20, text width=5em, text badly centered, minimum height=4em, inner sep=0pt, aspect=2]
\tikzstyle{block} = [rectangle, draw,fill=blue!20, text width=5em, text centered, minimum height=4em, rounded corners]
\tikzstyle{cloud} = [ellipse, draw,fill=red!20, text width=5em, text centered, minimum height=4em]
\tikzstyle{line} = [draw, -latex']
\tikzstyle{blockR} = [rectangle, draw, fill=red!20, text centered, minimum height=4em, rounded corners, minimum height=0.75cm]
\tikzstyle{blockB} = [rectangle, draw, fill=blue!20, text centered, minimum height=4em, rounded corners, minimum height=0.75cm]
\tikzstyle{blockG} = [rectangle, draw, fill=green!20, text centered, minimum height=4em, rounded corners, minimum height=0.75cm]
\tikzstyle{blockY} = [rectangle, draw, fill=yellow!20, text centered, minimum height=4em, rounded corners, minimum height=0.75cm]
\tikzstyle{blockW} = [rectangle, draw, fill=white!20, text centered, minimum height=4em, rounded corners, minimum height=0.75cm]
\tikzstyle{blockK} = [rectangle, draw, fill=gray!20, text centered, minimum height=4em, rounded corners, minimum height=0.75cm]
\tikzstyle{mblockB} = [rectangle, draw, fill=blue!20, text centered, minimum height=4em, double,rounded corners, minimum height=0.75cm]
\tikzstyle{mblockW} = [rectangle, draw, fill=white!20, text centered, minimum height=4em, double,rounded corners, minimum height=0.75cm]
\tikzstyle{mblockR} = [rectangle, draw, fill=red!20, text centered, minimum height=4em, double,rounded corners, minimum height=0.75cm]
\tikzstyle{mblockG} = [rectangle, draw, fill=green!20, text centered, minimum height=4em, double,rounded corners, minimum height=0.75cm]

\tikzstyle{circleW} = [circle, draw, fill=white!20, text centered, minimum height=4em, rounded corners, minimum height=0.75cm]

\begin{document}
\sloppy
\hbadness=5000

\pagestyle{empty}

\title{Synthesis of models 
for order-sorted first-order theories using linear algebra and constraint solving\thanks{Partially 
supported by the EU (FEDER), Spanish  
MINECO TIN 2013-45732-C4-1-P 
and GV PROMETEOII/2015/013.} }
\author{Salvador Lucas
\institute{DSIC,  Universitat Polit\`ecnica de Val\`encia, Spain\\ \url{http://users.dsic.upv.es/~slucas/}}}

\def\titlerunning{Synthesis of models 
for order-sorted first-order theories using linear algebra and constraint solving}

\def\authorrunning{Salvador Lucas}

\maketitle

\begin{abstract}
Recent developments in termination analysis for declarative programs
emphasize the use of appropriate models for the
logical theory representing the program at stake as a generic approach to
prove termination of declarative programs.
In this setting, \emph{Order-Sorted First-Order Logic} provides a powerful framework to 
represent declarative programs. It also provides a target logic to obtain models for other
logics via transformations.
We investigate the \emph{automatic generation} of numerical models for order-sorted first-order
logics and its use in program analysis, in particular in termination analysis of declarative programs.
We use \emph{convex domains} 
to give domains to the different \emph{sorts} of an order-sorted signature; 
we interpret the \emph{ranked}
symbols of sorted signatures by means of appropriately adapted \emph{convex matrix interpretations}.
Such \emph{numerical} interpretations 
permit the use of existing algorithms and tools from linear algebra 
and arithmetic constraint solving to synthesize the models.
\end{abstract}
\noindent
{\bf Keywords:~} Linear algebra, Logical models, Order-sorted first-order logic, Program Termination.

\section{Introduction}
\label{sec:intro}

In the \emph{logical approach to programming},
(declarative) programs are \emph{theories of a given logic $\cL$} 
and computation is \emph{deduction in the inference system} of $\cL$ \cite{Meseguer_GeneralLogics_LC87}.
The corresponding notion of \emph{termination} of declarative programs 
is the absence of infinite trees in any proof of a computation
\cite{LucMarMes_OpTermCTRSs_IPL05}.
Recently, a framework to \emph{prove} termination of declarative programs
has been developed \cite{LucMes_ProvingOperationalTerminationOfDeclarativeProgramsInGeneralLogics_PPDP14}.
In this framework, we obtain the \emph{proof jumps} associated to the inference system $\cI(\cS)$ which is
derived from the logic $\cL$ which is used to describe the program $\cS$.
Proof jumps are structures $\proofJump{B_1,\ldots,B_n}{A}$ where $n>0$ and $A,B_1,\ldots,B_n$ are 
\emph{formulas} in the inference rules $\frac{B_1\cdots B_n\cdots B_{n+p}}{A}$ in $\cI(\cS)$
(for $p\geq 0$).
Proof jumps are used to capture (infinite) paths in a proof tree $T$ 
using the rules in $\cI(\cS)$
so that there is a \emph{jump} from an instance $\sigma(A)$ of $A$ to an instance $\sigma(B_n)$ of $B_n$ 
provided that the corresponding intances of $B_1,\ldots,B_{n-1}$ were \emph{proved}, i.e., $\cS\vdash\sigma(B_i)$
for all $i$, $1\leq i<n$.
A set of proof jumps $\tau$ is an \emph{OT problem}. 
The \emph{initial} OT problem $\tau_I$ consists of all proof jumps for $\cI(\cS)$. 
Then, we apply an incremental proof methodology  which \emph{simplifies}  OT problems $\tau$
in a divide-and-conquer style to eventually prove (or disprove)
 termination of $\cS$. 
In particular, proof jumps $\psi:\proofJump{B_1,\ldots,B_n}{A}$
can be  \emph{removed} from an OT problem $\tau$ by using
\emph{well-founded relations} $\sqsupset$ as follows:
if, for all substitutions
$\sigma$, whenever $\cS\vdash\sigma(B_i)$ holds for all $i$, $1\leq i<n$, we have that 
$\sigma(A)\sqsupset\sigma(B_n)$ holds, then we can \emph{remove} $\psi$ from $\tau$.
In \cite{LucMes_ProvingOperationalTerminationOfDeclarativeProgramsInGeneralLogics_PPDP14} 
we show that \emph{logical models} are useful for this purpose. 
Any model $\SModel$ of $\cS$ satisfies the provable formulas, i.e., if $\cS\vdash\sigma(B_i)$ holds, 
then,
$\SModel\models\sigma(B_i)$ holds.
The point is using this fact to \emph{define} the well-founded relation $\sqsupset$.
This idea is  developed in \cite{LucMes_ModelsForLogicsAndConditionalConstraintsInAutomatedProofsOfTermination_AISC14}
for a systematic treatment of proofs of termination using logical models.

A sufficiently general and expressive framework to represent declarative programs, semantics of programming languages,
and program properties is \emph{Order-Sorted First-Order Logic} (OS-FOL), where the signature consists of 
a set $S$ of \emph{sorts} (i.e., names representing sets of values) 
which are ordered by  a \emph{subsort relation}
$\leq$ meaning subset inclusion, and 
 sets $\SSymbols_{w,s}$ and $\SPredicates_{w}$ 
of function and predicate symbols, where $s\in S$ and $w$ is a sequence $s_1\cdots s_k$ of sorts from $S$
\cite{GogMes_ModelsAndEqualityForLogicalProgramming_TAPSOFT87}. 
For instance, in our \emph{running example}  (Section \ref{SecRunningExample}) we develop
a termination analysis for an \emph{Order-Sorted Term Rewriting System} (OS-TRS) \cite{GogMes_ModelsAndEqualityForLogicalProgramming_TAPSOFT87,%
GogMes_OrderSortedAlgebraI_TCS92},
viewed as a particular case of OS-FOL theory with predicate symbols $\to$ and $\to^*$
describing the one-step rewrite relation $\to$ and the zero-or-more-steps relation $\to^*$,
see Figure \ref{fig:inferenceOSTRS}.
\begin{figure}[t]
\footnotesize
\begin{center}
\begin{tabular}{|@{\quad}l@{\quad}c@{\quad}l@{\quad}c@{\quad}|}
\hline
(Rf) & $\bigfrac{}{t \rightarrow^{\ast} t}$
& 
(T) & $\bigfrac{t \rightarrow t' \qquad t'
  \rightarrow^{*} u}{t  
\rightarrow^{*} u}$  
\\[0.3cm]
(C) & 
$\bigfrac{t_i \rightarrow t'_i}
{f(t_{1},\ldots,t_{i},\ldots,t_{k}){}\rightarrow{}f(t_{1},\ldots,t'_{i},\ldots,t_{k})}$ 
& 
(Re) &
$\bigfrac{}{
\ell  \to r }
$ 
\\ & where $f\in\Sigma_{w,s}$, $w=s_1,\ldots,s_k$, and $1\leq i\leq k$
& & where $\ell \rightarrow r \in\cR$ 
\\
\hline
\end{tabular}
\end{center}
\caption{Inference rules for Order-Sorted TRSs $\cR$}
\label{fig:inferenceOSTRS}
\end{figure}

In this paper we consider the automatic generation of models for OS-FOL theories.
This can be used in program analysis, in particular to mechanize the termination analysis of declarative programs as
explained above.
\emph{Semantic structures} $\SStructure'$ \cite{Hodges_AShorterModelTheory_1997} leading to \emph{decidable theories} $\mathit{Th}(\SStructure')$
\cite{Rabin_DecidableTheories_1977} can be used to provide an effective way to \emph{find} 
logical models $\SStructure$ for a program or specification $\cS$.
This is often possible by using \emph{theory transformations} $\kappa$ from the language of $\cS$ into
the language of  $\mathit{Th}(\SStructure')$ to obtain a set of sentences $\cS'=\kappa(\cS)$ which is then \emph{decidable}.
We formalize this view by extending the notion of \emph{derived algebra} 
\cite{GogThaWag_AnInitialAlgebraApproachToSpecificationCorrectnessAndImplementationOfADTs_1978} to logical structures.
Targeted languages usually involve symbols (e.g., $+$, $\times$,\ldots) with an
\emph{intended meaning} in the structures $\SStructure'$ that define the decidable theory $\mathit{Th}(\SStructure')$.
We also show how to transform an OS-FOL theory $\cS$ into a derived \emph{parametric}
theory $\cS^\sharp$ of linear arithmetic where appropriate algorithms and constraint solving techniques can
be used to give value to the parameters thus \emph{synthesizing} a \emph{model} of $\cS$.
The \emph{convex domains} introduced in \cite{LucMes_ModelsForLogicsAndConditionalConstraintsInAutomatedProofsOfTermination_AISC14} provide appropriate means for this.
They can be used to define bounded and unbounded domains for the sorts in the OS signature.
Indeed, the use of different (in particular \emph{bounded}) domains for some sorts is essential to
obtain a simple model which can be used to prove termination of our running example.

Section 
\ref{SecOrderSortedFirstOrderLogic} 
summarizes the basics of OS-FOL. 
Section \ref{SecDerivedModel} develops the notion of \emph{derived model} and shows how to use it to
deal with our running examples.
Section \ref{SecConstraintSolvingApproachToAnalyses} describes our automation approach using linear
algebra techniques and constraint solving.
Section \ref{SecOrderSortedStructuresWithConvexDomains} explains the generation of 
OS-FOL structures based on 
the \emph{convex domains} and \emph{convex matrix interpretations} introduced in \cite{LucMes_ModelsForLogicsAndConditionalConstraintsInAutomatedProofsOfTermination_AISC14}
Section \ref{SecAutomaticTreatmentOfTheRunningExample} 
shows how to apply the technique to obtain an \emph{automatic} solution to our case study.
Section \ref{sec:concl} concludes.

\subsection{Running example: termination of an order-sorted rewrite system}\label{SecRunningExample}

The OS-TRS $\pr{ToyamaOS}$ in Figure \ref{FigOSFOLTheoryForToyamaOS} 
is based on  
Toyama's example 
\cite{Toyama_CountExTermDirectSumTRSs_IPL87}.
It is given as a (hopefully self-explanatory) module of \Maude\ \cite{ClavelEtAl_MaudeBook_2007}.
\begin{figure}[t]
{\footnotesize
\begin{center}
\begin{minipage}{4cm}
\begin{verbatim}
mod ToyamaOS is
  sorts S S1 S2 .
  subsort S2 < S1 .
  op 0 : -> S2 .
  op 1 : -> S1 .
  op f : S1 S1 S1 -> S .
  op g : S1 S1 -> S1 .
  var x : S2 .
  vars  y z : S1 .
  rl f(0,1,x) => f(x,x,x) .
  rl g(y,z) => y .
  rl g(y,z) => z .
endm
\end{verbatim}
\end{minipage}
\begin{minipage}{10cm}
\begin{eqnarray}
& \forall t:\pr{S}~(t \to^* t)\label{ExToyamaOS_sentence1}\\
& \forall t:\pr{S1}~(t \to^* t)\label{ExToyamaOS_sentence2}\\\
& \forall t,t',u:\pr{S}~(t\to t'\wedge t' \to^* u\Rightarrow t\to^* u)\label{ExToyamaOS_sentence3}\\
& \forall t,t',u:\pr{S1}~(t\to t'\wedge t' \to^*u\Rightarrow t\to^* u)\label{ExToyamaOS_sentence4}\\
& \forall t_1,t'_1,t_2,t_3:\pr{S1}~(t_1\to t'_1 \Rightarrow \pr{f}(t_1,t_2,t_3)\to \pr{f}(t'_1,t_2,t_3))\label{ExToyamaOS_sentence5}\\
& \forall t_1,t_2,t'_2,t_3:\pr{S1}~(t_2\to t'_2 \Rightarrow \pr{f}(t_1,t_2,t_3)\to \pr{f}(t_1,t'_2,t_3))\label{ExToyamaOS_sentence6}\\
& \forall t_1,t_2,t_3,t'_3:\pr{S1}~(t_3\to t'_3 \Rightarrow \pr{f}(t_1,t_2,t_3)\to \pr{f}(t_1,t_2,t'_3))\label{ExToyamaOS_sentence7}\\
& \forall t_1,t'_1,t_2:\pr{S1}~(t_1\to t'_1 \Rightarrow \pr{g}(t_1,t_2)\to \pr{g}(t'_1,t_2))\label{ExToyamaOS_sentence8}\\
& \forall t_1,t_2,t'_2,t_3:\pr{S1}~(t_2\to t'_2 \Rightarrow \pr{g}(t_1,t_2)\to \pr{g}(t_1,t'_2))\label{ExToyamaOS_sentence9}\\
& \forall x:\pr{S2}~(\pr{f}(\pr{0},\pr{1},x)\to \pr{f}(x,x,x))\label{ExToyamaOS_sentence10}\\
& \forall x,y:\pr{S1}~(\pr{g}(x,y)\to x)\label{ExToyamaOS_sentence11}\\
& \forall x,y:\pr{S1}~(\pr{g}(x,y)\to y)\label{ExToyamaOS_sentence12}
\end{eqnarray}

\end{minipage}
\end{center}}
\caption{Order-sorted version of Toyama's example and its associated Order-Sorted First-Order Theory}
\label{FigOSFOLTheoryForToyamaOS}
\end{figure}
The unsorted version of this module is nonterminating \cite{Toyama_CountExTermDirectSumTRSs_IPL87}.
Furthermore, if $\pr{S1}$ and $\pr{S2}$ are confused into a single sort 
then $\pr{ToyamaOS}$ is nonterminating too:
\[\pr{f(\ul{g(0,1)},g(0,1),g(0,1))}\to\pr{f(0,\ul{g(0,1)},g(0,1))}\to\ul{\pr{f(0,1,g(0,1))}}\to\pr{f(\ul{g(0,1)},g(0,1),g(0,1))}\to\cdots\]
But with all sort information we can it prove it \emph{terminating}. For instance, variable $\pr{x}$ (of sort
$\pr{S2}$) cannot
be bound to terms of sort $\pr{S1}$ which is a supersort of $\pr{S2}$.
Thus, the third step, which requires a binding $\pr{x}\mapsto\pr{g(0,1)}$, is not possible because
the sort of $\pr{g(0,1)}$ is $\pr{S1}$.
Thus the infinite sequence is not possible.

The \emph{order-sorted first-order theory} for the OS-TRS is also shown in
Figure \ref{FigOSFOLTheoryForToyamaOS}.
It is obtained by specializing the inference rules in Figure \ref{fig:inferenceOSTRS}.
Sentences in Figure \ref{FigOSFOLTheoryForToyamaOS}
make explicit the implicit quantification of the inference rules by 
taking into account the sorts in the signature and the subsort ordering.
In particular, the only quantification over $\pr{S2}$ occurs in (\ref{ExToyamaOS_sentence10}).
It turns out that such a quantification is crucial to obtain a simple proof of termination.
In order to prove termination of this OS-TRS  we need to find a model $\SModel$ for the theory in Figure \ref{FigOSFOLTheoryForToyamaOS} such 
that $\to$ is interpreted as a \emph{well-founded relation} $>$.
Although we do not have space to further justify this claim, it easily follows from the theory in \cite{LucMes_ProvingOperationalTerminationOfDeclarativeProgramsInGeneralLogics_PPDP14}. 

\section{Order-Sorted First-Order Logic}\label{SecOrderSortedFirstOrderLogic}

\paragraph*{\bf Sorts and Order-Sorted Signatures.}
Given a set of \emph{sorts} $S$, a many-sorted signature
is an $S^{\ast} \times S$-indexed family of sets $\Sigma =
\{\Sigma_{w,s}\}_{(w,s) \in S^{\ast} \times S}$ containing
\emph{function symbols} with a given string of argument sorts and a result
sort. If $f \in \Sigma_{s_{1}\cdots s_{n},s}$, then we display
$f$ as $f: s_{1}\cdots s_{n} \to s$. 
This is called a \emph{rank} declaration for symbol $f$.
Constant symbols $c$ (taking no argument) have rank declaration $c:\lambda\to s$ for some sort $s$ (where $\lambda$ 
denotes the \emph{empty} sequence).
An order-sorted signature $(S,\leq,\SSymbols)$
consists of a poset of sorts $(S,\leq)$ together with a many-sorted signature $(S,\SSymbols)$.
The \emph{connected components} of $(S,\leq)$ are the equivalence classes
$[s]$ corresponding to the least equivalence relation $\equiv_\leq$ containing $\leq$.
We extend the order $\leq$ on $S$ to strings of equal length in $S^*$ by  $s_1\cdots s_n\leq s'_1\cdots s'_n$ iff $s_i\leq s'_i$
for all $i$, $1\leq i\leq n$.
Symbols $f$ can be \emph{subsort-overloaded}, i.e., they can have
several rank declarations related in the $\leq$ ordering \cite{GogMes_OrderSortedAlgebraI_TCS92}.
Constant symbols, however, have only one rank declaration.
Besides, the following \emph{monotonicity condition} must be satisfied: 
$f\in\SSymbols_{w_1,s_1}\cap\SSymbols_{w_2,s_2}$ and $w_1\leq w_2$ imply $s_1\leq s_2$. 
To avoid ambiguous terms, we assume that $\SSymbols$ is \emph{sensible}, meaning that
if $f:s_1\cdots s_n\to s$ and $f:s'_1\cdots s'_n\to s'$ are such that $[s_i]=[s'_i]$, $1\leq i\leq n$,
then $[s]=[s']$.
Throughout this paper, $\SSymbols$ will always be assumed \emph{sensible}.
An order-sorted signature $\SSymbols$ is \emph{regular} iff given $w_0\leq w_1$ in $S^*$ and $f\in\SSymbols_{w_1,s_1}$, there is a least $(w,s)\in S^*\times S$ such that $f\in\SSymbols_{w,s}$ and $w_0\leq w$.
If, in addition, each connected component $[s]$ of the sort poset has a top element $\top_{[s]}\in [s]$, then the regular signature
is called \emph{coherent}.

Given an $S$-sorted set $\Variables=\{\Variables_s\mid s\in S\}$ of \emph{mutually disjoint}
sets of variables (which are also disjoint from the signature $\SSymbols$), 
the set $\STerms_s$ of terms of sort $s$ is the least set such that
(i) $\Variables_{s}\subseteq\STerms_s$,
(ii) 
If $s'\leq s$, then $\STerms_{s'}\subseteq\STerms_{s}$; and 
(iii) for each $f: s_{1}\cdots s_{n} \to s$ 
and $t_i\in\STerms_{s_i}$, $1\leq i\leq n$,  $f(t_1,\ldots,t_n)\in\STerms_s$. 
If $\Variables=\emptyset$, we write $\GSTerms$ rather than $\STermsOn{\SSymbols}{\emptyset}$ for the set of \emph{ground} terms.
Terms with variables can also be seen as a special case of ground terms of the \emph{extended} signature $\SSymbols(\Variables)$
where variables are considered as \emph{constant} symbols of the apporpriate sort, i.e., $\SSymbols(\Variables)_{\lambda,s}=\SSymbols_{\lambda,s}\cup\Variables_s$.

\begin{example}\label{ExToyamaOS_Signature}
The order-sorted signature $(S,\leq,\SSymbols)$ for program $\pr{ToyamaOS}$ consists of the
following components:
\begin{enumerate}
\item\label{ExToyamaOS_Signature_Sorts}  \emph{Set of sorts} $S=\{\pr{S},\pr{S1},\pr{S2}\}$.
\item\label{ExToyamaOS_Signature_OrderingOnSorts}
The \emph{subsort relation} is the least ordering $\leq$ on $S$ satisfying 
$\pr{S2}\leq\pr{S1}$.
\item\label{ExToyamaOS_Signature_ConnectedComponents}
Thus, $(S,\leq)$ (or $S/_{\equiv_\leq}$) consists of two \emph{connected components}: 
$[\pr{S}]=\{\pr{S}\}$ and
$[\pr{S1}]=\{\pr{S2},\pr{S1}\}$.
\item\label{ExToyamaOS_Signature_TopSorts}
Note that $\pr{S}$ is the \emph{top sort} $\top_{[\pr{S}]}$ of  $[\pr{S}]$, and 
$\pr{S1}$ is the top sort $\top_{[\pr{S1}]}$ of $[\pr{S1}]$.
\item\label{ExToyamaOS_Signature_Symbols}
The signature is $\SSymbols=\SSymbols_{\fpr{S1}}\cup\SSymbols_{\fpr{S2}}\cup\SSymbols_{\fpr{S1}\:\fpr{S1},\fpr{S1}}\cup\SSymbols_{\fpr{S1}\:\fpr{S1}\:\fpr{S1},\fpr{S}}$, with
$\SSymbols_{\fpr{S1}}=\{\pr{1}\}$,
$\SSymbols_{\fpr{S2}}=\{\pr{0}\}$,
$\SSymbols_{\fpr{S1}\:\fpr{S1},\fpr{S1}}=\{\pr{g}\}$, and
$\SSymbols_{\fpr{S1}\:\fpr{S1}\:\fpr{S1},\fpr{S}}=\{\pr{f}\}$.
\item\label{ExToyamaOS_Signature_RegularityAndCoherence}
There is no overloaded function symbol, i.e., $\SSymbols$ is trivially \emph{regular}.
Furthermore, since every connected component has a top sort (see item \ref{ExToyamaOS_Signature_TopSorts}), 
$(S,\leq,\SSymbols)$ is a \emph{coherent} signature.

\end{enumerate}
The set of variables is
$\Variables=
\Variables_{\fpr{S1}}\cup\Variables_{\fpr{S2}}$,
with 
$\Variables_{\fpr{S1}}=\{\pr{y},\pr{z}\}$, and
$\Variables_{\fpr{S2}}=\{\pr{x}\}$.
\end{example}
The assumption that $\SSymbols$ is sensible ensures that if $[s]\neq [s']$, then 
$\STerms_{[s]}\cap\STerms_{[s']}=\emptyset$.
The set $\STerms$ of order-sorted terms is $\STerms=\cup_{s\in S}\STerms_s$.
An element of any set $\STerms_s$ is called a \emph{well-formed} term.

\paragraph*{\bf Order-Sorted Algebras.}
Given a many-sorted signature  $(S,\SSymbols)$, an  $(S,\SSymbols)$-algebra $\SAlgebra$
(or just a $\SSymbols$-algebra, if $S$ is clear from the context)  is a family $\{\SemDomain_s\mid s\in S\}$ of sets called the \emph{carriers}
or \emph{domains}
of $\SAlgebra$ together with a function $f^\SAlgebra_{w,s}\in\SemDomain_w\to\SemDomain_s$ for each $f\in\SSymbols_{w,s}$ where $\SemDomain_w=\SemDomain_{s_1}\times\cdots\times \SemDomain_{s_n}$ if $w=s_1\cdots s_n$, and $\SemDomain_w$ is a one 
point set when $w=\lambda$. 
Given an order-sorted signature $(S,\leq,\SSymbols)$, an $(S,\leq,\SSymbols)$-algebra (or $\SSymbols$-algebra if $(S,\leq)$ is 
clear from the context) is an $(S,\SSymbols)$-algebra such that
\begin{enumerate}
\item If $s,s'\in S$ are such that $s\leq s'$, then $\SemDomain_s\subseteq \SemDomain_{s'}$, and
\item\label{ItemConditionsOverloadedFunction_EqualOverSubSorts} 
If $f\in\SSymbols_{w_1,s_1}\cap\SSymbols_{w_2,s_2}$ and $w_1\leq w_2$, then $f^\SAlgebra_{w_1,s_1}\in\SemDomain_{w_1}\to A_{s_1}$
equals $f^\SAlgebra_{w_2,s_2}\in\SemDomain_{w_2}\to A_{s_2}$ on $\SemDomain_{w_1}$.
\end{enumerate}
\begin{remark}\label{RemOverloadedSymbolsAreGivenDifferentFunctions}
Note that overloaded symbols $f$ may be given \emph{different} functions $f^\SAlgebra_{w_1,s_1},\ldots,f^\SAlgebra_{w_n,s_n}$ 
depending on the specific ranks $w_1\to s_1,\ldots,w_n\to s_n$ of the overload for symbol $f$.
Of course, such functions must still fulfill condition \ref{ItemConditionsOverloadedFunction_EqualOverSubSorts} above.
\end{remark}
With regard to many sorted signatures and algebras, an 
$(S,\SSymbols)$-homomorphism between $(S,\SSymbols)$-algebras $\SAlgebra$ and $\SAlgebra'$
is an $S$-sorted function $h=\{h_s:\SemDomain_s\to\SemDomain'_s\mid s\in S \}$ such that
for each $f\in\SSymbols_{w,s}$ with $w=s_1,\ldots,s_k$, 
$h_s(f^\SAlgebra_{w,s}(a_1,\ldots,a_k))=f^{\SAlgebra'}_{w,s}(h_{s_1}(a_1),\ldots,h_{s_k}(a_k))$.
If $w=\lambda$, we have $h_s(f^\SAlgebra)=f^{\SAlgebra'}$.
Now, for the order-sorted case, an $(S,\leq,\SSymbols)$-homomorphism $h:\SAlgebra\to\SAlgebra'$
between $(S,\leq,\SSymbols)$-algebras $\SAlgebra$ and $\SAlgebra'$ is an $(S,\SSymbols)$-homomorphism
that satisfies the following additional condition: if $s\leq s'$ and $a\in\SemDomain_s$, then $h_s(a)=h_{s'}(a)$.

The family of \emph{domains} $\{\STerms_s\}_{s\in S}$ together with the operations $f:(t_1,\ldots,t_n)\mapsto f(t_1,\ldots,t_n)$
define an order-sorted $\SSymbols$-algebra called the \emph{free} algebra on $\Variables$
and denoted $\STerms$.
When $\Variables=\emptyset$, $\GSTerms=\STermsOn{\SSymbols}{\emptyset}$ denotes the \emph{initial}
$\SSymbols$-algebra, i.e., an algebra having a unique homomorphism $h_\SAlgebra:\GSTerms\to\SAlgebra$ to each 
$\SSymbols$-algebra $\SAlgebra$.
Similarly, $\STerms$ (itself a $\SSymbols$-algebra) is initial in the class of all $\SSymbols(\Variables)$-algebras.

\paragraph*{\bf Predicates and connectives.}

Following \cite{GogMes_ModelsAndEqualityForLogicalProgramming_TAPSOFT87},  
an order-sorted signature \emph{with predicates} 
is a quadruple ${(S,\leq,\SSymbols,\SPredicates)}$
such that $(S,\leq,\SSymbols)$ is an coherent 
order-sorted signature, and $\Pi=\{\Pi_w\mid w\in S^+\}$
is a family of \emph{predicate symbols} $P$, $Q$, \ldots
We write $P:w$ for $P\in\Pi_w$.
Overloading is also allowed on predicates
with the following conditions \cite[Definition 11]{GogMes_ModelsAndEqualityForLogicalProgramming_TAPSOFT87}:
\begin{enumerate}
\item There is an equality predicate symbol $=\:\in\Pi_{ss}$ iff $s$ is the top of a connected component
of the sort poset $S$.
\item \emph{Regularity}: For each $w_0$ such that there is $P\in\Pi_{w_1}$ with $w_0\leq w_1$, there is a least $w$
such that $P\in\Pi_w$ and $w_0\leq w$.
\end{enumerate}
We often write $\SSymbols,\SPredicates$ instead of $(S,\leq,\SSymbols,\SPredicates)$ if
$S$ and $\leq$ are clear from the context.
The formulas $\varphi$ of an order-sorted signature with predicates $\SSymbols,\SPredicates$
are built up from atoms $P(t_1,\ldots,t_n)$ with $P\in\Pi_{w}$ and $t_1,\ldots,t_n\in\STerms_{w}$, 
logic connectives (e.g., $\wedge$, $\neg$)
and quantifiers ($\forall$) as follows: 
(i) if $P\in\SPredicates_{w}$, $w=s_1\cdots s_n$, and $t_i\in\STerms_{s_i}$ for all $i$, $1\leq i\leq n$,
then $P(t_1,\ldots,t_n)\in\mathit{Form}_{\SSymbols,\SPredicates}$.
(ii) if $\phi\in\mathit{Form}_{\SSymbols,\SPredicates}$, then
$\neg\phi\in\mathit{Form}_{\SSymbols,\SPredicates}$;
(iii) if $\phi,\phi'\in\mathit{Form}_{\SSymbols,\SPredicates}$, then $\phi\wedge\phi'\in\mathit{Form}_{\SSymbols,\SPredicates}$;
(iv) if $s\in S$,  $x\in\Variables_s$, and $\phi\in\mathit{Form}_{\SSymbols,\SPredicates}$, then 
$(\forall x:s)\phi\in\mathit{Form}_{\SSymbols,\SPredicates}$.
As usual, we can consider formulas involving other logic connectives and quantifiers (e.g., $\vee$, $\Rightarrow$, 
$\Leftrightarrow$, $\exists$,...) by using their standard definitions in terms of $\wedge$, $\neg$, $\forall$.
A closed formula, i.e., whose variables are all universally or existentially quantified, is called a \emph{sentence}.

\begin{remark}
In order to define an order-sorted signature with predicates that can be used to reason about rewritings with OS-TRSs,
we have to provide (at least) as many overloads for the computational relation $\to^*$ as connected
component $[s]$ in $S/_{\equiv_{\leq}}$:
due to axiom (Rf), OS-TRSs are expected to rewrite with $\to^*$ any of the classes $\STerms_{[s]}$ for every connected
component $[s]$.
By coherence of the signature, we can just let $\to^*\in\Pi_{\top_{[s]}\:\top_{[s]}}$ for all $s\in S$.
Then, rule (T) requires a corresponding overload for $\to$ as well.
By coherence of the signature, we can just let $\Pi_{\top_{[s]}\:\top_{[s]}}=\{\to,\to^*\}$ for all $s\in S$.
This will be compatible with any possible instance of rule (Re) because terms $\ell$ and $r$ in 
rewrite rules $\ell\to r$ of OS-TRSs must be terms belonging to $\STerms_{[s]}$ for some $s\in S$.
By coherence, we know that $\ell,r\in\STerms_{\top_{[s]}}$ for some $s\in S$.
\end{remark}
\begin{example}\label{ExToyamaOS_SignatureWithPredicates}
The order-sorted signature $(S,\leq,\SSymbols)$ described in Example \ref{ExToyamaOS_Signature} is
extended into a order-sorted signature with predicates $(S,\leq,\SSymbols,\Pi)$ where $\Pi=\Pi_{\fpr{S}\:\fpr{S}}\cup\Pi_{\fpr{S1}\:\fpr{S1}}$
for $\Pi_{\fpr{S}\:\fpr{S}}=\Pi_{\fpr{S1}\:\fpr{S1}}=\{\to,\to^*\}$, which are the only nonempty sets of predicate symbols. They
satisfy the regularity condition.
\end{example}

\paragraph*{\bf Theories, specifications and programs.}
A \emph{theory} $\cS$ of $\SSymbols,\SPredicates$ is a set of formulas,
$\cS\subseteq\mathit{Form}_{\SSymbols,\SPredicates}$, and its \emph{theorems}  are the formulas 
$\varphi\in\mathit{Form}_{\SSymbols,\SPredicates}$
for which we can derive a proof using an appropriate inference system $\cI(\cL)$ of a logic $\cL$ in the usual way
(written $\cS\vdash\varphi$).
Given a logic $\cL$ describing computations in a (declarative) programming language, 
programs are viewed as \emph{theories} $\cS$ of $\cL$.

\begin{example}\label{ExOSToyama_Sentences}
In the logic of OS-TRSs, with binary (overloaded) \emph{predicates} $\to$ and $\to^*$, 
the theory for an
OS-TRS $\cR=(S,\leq,\SSymbols,R)$ with set of rules $R$ (for instance, our running example) 
is obtained from  the \emph{schematic} inference rules in Figure \ref{fig:inferenceOSTRS} after 
\emph{specializing} them  
as $(\mathit{C})_{f,i}$ for each $f\in\Symbols$ and $i$, $1\leq i\leq ar(f)$
and $(\mathit{Re})_\rho$ for all $\rho:\ell\to r\in R$.
Then,  inference rules $\frac{B_1,\ldots,B_n}{A}$ become \emph{implications} 
$B_1\wedge\cdots\wedge B_n\Rightarrow A$.
For instance, with regard to the sentences for \verb$ToyamaOS$ in Figure \ref{FigOSFOLTheoryForToyamaOS}:
\begin{itemize}
\item Sentences (\ref{ExToyamaOS_sentence1}) and (\ref{ExToyamaOS_sentence2}) specialize (Rf) in Figure
\ref{fig:inferenceOSTRS} for the two overloads of $\to^*$ in $\Pi_{\pr{S}\:\pr{S}}$ and $\Pi_{\pr{S1}\:\pr{S1}}$, respectively.
\item Sentences (\ref{ExToyamaOS_sentence3}) and (\ref{ExToyamaOS_sentence4}) specialize (T)
 for the overloads of $\to^*$ and $\to$ in $\Pi_{\pr{S}\:\pr{S}}$ and $\Pi_{\pr{S1}\:\pr{S1}}$, respectively.
\item Sentences (\ref{ExToyamaOS_sentence5}), (\ref{ExToyamaOS_sentence6}), and (\ref{ExToyamaOS_sentence7}) specialize (C)
 for symbol $\pr{f}$ using the appropriate overloads of $\to$ in $\Pi_{\pr{S}\:\pr{S}}$ and $\Pi_{\pr{S1}\:\pr{S1}}$ according to the rank of
 $\pr{f}$. Similarly, (\ref{ExToyamaOS_sentence8}) and (\ref{ExToyamaOS_sentence9}) specialize (C)
 for symbol $\pr{g}$.
 \item Sentences (\ref{ExToyamaOS_sentence10}), (\ref{ExToyamaOS_sentence11}), and (\ref{ExToyamaOS_sentence12}) specialize
 (Re) for each rewrite rule in \verb$ToyamaOS$.
\end{itemize}
Note that, according to the variable declaration for $\pr{x}$ in  \verb$ToyamaOS$, 
in sentence (\ref{ExToyamaOS_sentence10}) variable $x$ ranges on values of sort $\pr{S2}$ only.
\end{example}

\paragraph*{\bf Structures, Satisfaction, Models.}
Given an order-sorted signature with predicates ${(S,\leq,\SSymbols,\SPredicates)}$,  an $(S,\leq,\SSymbols,\SPredicates)$-\emph{structure}\footnote{As in \cite{Hodges_AShorterModelTheory_1997}, we use 
`structure' and reserve the word `model' to refer those structures satisfying a given theory.}  
(or just a $\SSymbols,\SPredicates$-structure)
is an order-sorted $(S,\leq,\SSymbols)$-algebra $\SStructure$ together with an assignment
to each $P\in\Pi_w$ of a subset $P^\SStructure_w\subseteq\SStructure_{w}$
 such that \cite{GogMes_ModelsAndEqualityForLogicalProgramming_TAPSOFT87}:
 (i) for $P$ the identity predicate $\_=\_:ss$, the assignment is the identity relation, i.e., $(=)_{\StructDomain}=\{(a,a)\mid a\in \StructDomain_s\}$; and
 (ii) whenever $P:w_1$ and $P:w_2$ and $w_1\leq w_2$, then $P^{\StructDomain}_{w_1}=\StructDomain_{w_1}\cap P^\StructDomain_{w_2}$.

Let $(S,\leq,\SSymbols,\SPredicates)$ be an order-sorted signature with predicates and $\SStructure,\SStructure'$ be
$(S,\leq,\SSymbols,\SPredicates)$-structures.
Then, an \emph{$(S,\leq,\SSymbols,\SPredicates)$-homomorphism} $h:\SStructure\to\SStructure'$ is 
an $(S,\leq,\SSymbols)$-homomorphism such that, for each $P:w$ in $\Pi$, 
if $(a_1,\ldots,a_n)\in P^{\SStructure}_w$, then $h(a_1,\ldots,a_n)\in P^{\SStructure'}_w$. 

Given an $S$-sorted  \emph{valuation mapping} $\alpha:\Variables\to\SemDomain$, the evaluation mapping
$[\_]^\alpha_\SStructure:\STerms\to \SemDomain$ 
is the unique $(S,\leq,\SSymbols)$-homomorphism extending $\alpha$ \cite{GogMes_OrderSortedAlgebraI_TCS92}. 
Finally, $[\_]^\alpha_\SStructure:\mathit{Form}_{\SSymbols,\SPredicates}\to\mathit{Bool}$ is given by: 
\begin{enumerate}
\item $[P(t_1,\ldots,t_k)]^\alpha_\SStructure=\Ftrue$ for $P:w$ and terms $t_1,\ldots,t_k$ if and only if $([t_1]^\alpha_\SStructure,\ldots,[t_k]^\alpha_\SStructure)\in P^\ModelDomain_w$;
\item $[\neg\phi]^\alpha_\SStructure=\Ftrue$ if and only if $[\phi]^\alpha_\SStructure=\Ffalse$;
\item $[\phi\wedge\psi]^\alpha_\SStructure=\Ftrue$ if and only if $[\phi]^\alpha_\SStructure=\Ftrue$ and $[\psi]^\alpha_\SStructure=\Ftrue$;
\item $[(\forall x:s)\:\phi]^\alpha_\SStructure=\Ftrue$ if and only if for all $a\in\SemDomain_s$, $[\phi]^{\alpha[x\mapsto a]}_\SStructure=\Ftrue$;
\end{enumerate}
We say that $\SStructure$ \emph{satisfies} $\varphi\in\mathit{Form}_{\SSymbols,\SPredicates}$ if there is 
$\alpha\in\Variables\to\SemDomain$ such that $[\varphi]^\alpha_\SStructure=\Ftrue$.
If $[\varphi]^\alpha_\SStructure=\Ftrue$ for \emph{all} valuations $\alpha$, we write $\SStructure\models\varphi$
and say that $\SStructure$ is a \emph{model} of $\varphi$ \cite[page 12]{Hodges_AShorterModelTheory_1997}.
Initial valuations are not relevant for establishing the satisfiability of \emph{sentences};
thus, both notions coincide on them.
We say that $\SModel$ is \emph{a model of a set of sentences} $\cS\subseteq\mathit{Form}_{\SSymbols,\SPredicates}$ (written $\SModel\models\cS$)
if for all $\varphi\in\cS$, $\SModel\models\varphi$.
And, given a sentence $\varphi$, we write $\cS\models\varphi$ if and only if for \emph{all models} 
$\SModel$ of $\cS$, $\SModel\models\varphi$.
\emph{Sound} logics guarantee that every provable sentence $\varphi$ is true in \emph{every model} of $\cS$, i.e., $\cS\vdash\varphi$
implies $\cS\models\varphi$.

\section{Derived models}\label{SecDerivedModel}

By a \emph{decidable theory}  $T$ in a given
language (often a fragment of first-order logic) we mean one having a \emph{decision procedure} which can be used to
establish whether a given formula $\phi$ belongs to $T$ \cite{Rabin_DecidableTheories_1977}.
In some cases such theories can be presented as \emph{axiomatizations} of algebraic structures $\SStructure$
so that $T=Th(\SStructure)=\{\phi\mid \SStructure\models \phi\}$.
We often say that $\SStructure$ is the \emph{intended model} of $T$ \cite[page 32]{Hodges_AShorterModelTheory_1997}.

\begin{example}\label{ExPresburgerArithmeticIsADecidableTheory}
Presburger's arithmetic (or arithmetic without multiplication) can be seen as the set  of sentences of the language
$L_P=\{\pr{0},\pr{'},\pr{+},\pr{>}\}$ 
which are true in the standard interpretation $\cN$ of the natural numbers \cite[page 295]{BooBurJef_ComputabilityAndLogic_2002}. 
It is well-known that $P=Th(\cN)$ is decidable.
\end{example}
Assume that $(S',\leq',\SSymbols',\SPredicates')$ is an order-sorted signature with predicates 
and $\SStructure'$ is a $\SSymbols',\SPredicates'$-structure such that
 $T=Th(\SStructure')$ is \emph{decidable}.
 We can define an $(S,\leq,\SSymbols,\SPredicates)$-model  for $\cS\subseteq\mathsf{Form}_{\SSymbols,\SPredicates}$ 
by means of a map 
(theory transformation) 
$\kappa:\mathsf{Form}_{\SSymbols,\SPredicates}\to\mathsf{Form}_{\SSymbols',\SPredicates'}$.
If $\cS$ is finite, then, it is \emph{decidable} whether $\kappa(\cS)\subseteq T$.
If $\kappa(\cS)\subseteq T$, then $\SStructure'\models\kappa(\cS)$, i.e., the $\SSymbols',\SPredicates'$-structure
$\SStructure'$ is a model of $\kappa(\cS)$.
If we can define $\kappa$ on a purely syntactic basis, i.e., as homomorphic extensions of maps 
from the syntactic components $S$, $\SSymbols$, and $\SPredicates$ in $(S,\leq,\SSymbols,\SPredicates)$, 
then we are able to make $\SStructure'$ into a \emph{derived} $\SSymbols,\SPredicates$-structure $\SStructure$
so that $\SStructure$ is a  model of $\cS$, i.e., $\SStructure\models\cS$, as desired.
In the following, we further develop this methodology.

\subsection{Derived algebras and structures}

Appropriate $\SSymbols$-algebras can be obtained as \emph{derived algebras} if we first 
consider a \emph{new} signature $\SSymbols'$ 
of symbols  
with `intended' (often arithmetic) interpretations.

\begin{definition}[Derivor and Derived algebra]{\rm \cite[Definition 11]{GogThaWag_AnInitialAlgebraApproachToSpecificationCorrectnessAndImplementationOfADTs_1978}}
\label{DerivedAlgebra}
Let $\SSymbols=(S,\leq,\SSymbols)$ and $\SSymbols'=(S',\leq',\SSymbols')$ be order-sorted signatures.
A \emph{derivor} from $\SSymbols$ to $\SSymbols'$ is a monotone function 
$\tau:S\to S'$ (i.e., such that for all $s,s'\in S$, $s\leq s'$ implies\footnote{Monotonicity is \emph{not} required
in \cite{GogThaWag_AnInitialAlgebraApproachToSpecificationCorrectnessAndImplementationOfADTs_1978} where only many-sorted signatures are considered.} $\tau(s)\leq'\tau(s')$) 
and a family $d_{w,s}:\SSymbols_{w,s}\to(\GSTerms)_{\tau(w),\tau(s)}$, where $\tau(s_1,\ldots,s_k)=\tau(s_1),\ldots,\tau(s_k)$ and where $(\GSTerms)_{\tau(w),\tau(s)}$ denotes the set of all $\SSymbols'$-terms
using variables $\{y_1,\ldots,y_k\}$ with $y_i$ of sort $\tau(s_i)$.
Each operation symbol $f\in\SSymbols_{w,s}$ is expressed using a derived operation $d_{w,s}(f)$ of the appropriate arity.
We often use $d$ to denote a derivor $\langle\tau,d\rangle$.
Now, let $\SAlgebra'$ be an $\SSymbols'$-algebra. 
Then, the $d$-derived algebra $d\SAlgebra'$ of $\SAlgebra'$ is the $\SSymbols$-algebra with carriers 
$(d\SAlgebra')_s=\SAlgebra'_{\tau(s)}$ for all $s\in S$; and
mappings $f^{d\SAlgebra'}$ for each $f\in\SSymbols$ defined to be $(d(f))^{\SAlgebra'}$, the derived operator of the $\SSymbols'$-term $d(f)$.
\end{definition}
Note that $d$ in Definition \ref{DerivedAlgebra} is \emph{homomorphically extended} into a 
mapping $d:\STerms\to\STermsOn{\SSymbols'}{\Variables'}$.
\begin{example}\label{ExToyamaOS_DerivedAlgebra}
Let $(S,\leq,\SSymbols)$ as in Example \ref{ExToyamaOS_Signature}.
Let $S'=\{\mathsf{zero},\mathsf{nat}\}$ with subsort relation $\leq'$ 
given by
$\mathsf{zero}\leq'\mathsf{nat}$, and
$\SSymbols'=\SSymbols'_{\lambda,\mathsf{zero}}\cup\SSymbols'_{\lambda,\mathsf{one}}\cup\SSymbols'_{\mathsf{nat}^2\mathsf{nat}}$ 
where 
$\SSymbols_{\lambda,\mathsf{zero}}=\{\pr{0}\}$, 
$\SSymbols_{\lambda,\mathsf{nat}}=\{\pr{1}\}$, 
and $\SSymbols_{\mathsf{nat}^2\mathsf{nat}}=\{\pr{+}\}$.
We define 
a derivor from $(S,\leq,\SSymbols)$ to $(S',\leq',\SSymbols')$ by
$\tau(\pr{S})=\tau(\pr{S1})=\mathsf{nat}$ and
$\tau(\pr{S2})=\mathsf{zero}$; 
also, 
$d(\pr{0})=\pr{0}$, 
$d(\pr{1})=\pr{1}$,  
$d(\pr{f})=x\:\pr{+}\:y\:\pr{+}\:z$, and $d(g)=x\:\pr{+}\:y\:\pr{+}\:\pr{1}$.
Let $\SAlgebra'$ be the $(S',\leq',\SSymbols')$ algebra  given by
$\SAlgebra'_{\mathsf{zero}}=\{0\}$ and
$\SAlgebra'_{\mathsf{nat}}=\naturals$ 
together with the \emph{standard} interpretations for $0$, $1$, and $+$. 
The derived $(S,\leq,\SSymbols)$-algebra $\SAlgebra=d\SAlgebra'$ is  given by
$\SAlgebra_{\pr{S2}}=\SAlgebra'_{\mathsf{zero}}=\{0\}$  and
$\SAlgebra_{\pr{S}}=\SAlgebra_{\pr{S1}}=\SAlgebra'_{\mathsf{nat}}=\naturals$, 
together with the derived interpretations for each symbol in $\SSymbols$.
\end{example}
A slight generalization of Definition \ref{DerivedAlgebra} leads to the notion of 
\emph{derived structure}.

\begin{definition}[Derivor for signatures with predicates / Derived structure]
\label{DerivedStructure}
Let $\SSymbols=(S,\leq,\SSymbols,\SPredicates)$ and $\SSymbols'=(S',\leq',\SSymbols',\SPredicates')$ be order-sorted signatures with predicates and $\langle\tau,d\rangle$ be a derivor from $(S,\leq,\SSymbols)$ to 
$(S',\leq',\SSymbols')$.
We extend $d$ to predicate symbols by adding a component $d:\SPredicates\to\mathsf{Form}_{\SSymbols'\SPredicates'}$ such that for all $P\in\SPredicates_{w}$,
with $w=s_1\cdots s_n$, $d(P)$ is an \emph{atom} $P'(t'_1,\ldots,t'_m)$ with $P'\in\SPredicates'_{w'}$, 
and terms $t'_1,\ldots,t'_m\in\STermsOn{\SSymbols'}{\Variables}$ only use variables $\{y_1,\ldots,y_n\}$ with 
$y_i$ of sort $\tau(s_i)$.
In this new context we also call $\langle\tau,d\rangle$ a derivor.
Let $\SStructure'=(\SemDomain',\SSymbols_{\SStructure'},\Pi'_{\SStructure'})$ be an 
$(S',\leq',\SSymbols',\SPredicates')$-structure
and $\SAlgebra'_0=(\SemDomain',\SSymbols_{\SModel'})$ be the underlying $(S',\leq',\SSymbols')$-algebra. 
Then, the $\langle\tau,d\rangle$-derived structure $d\SStructure'$ of $\SStructure'$ is the 
$(S,\leq,\SSymbols,\SPredicates)$-structure
that consists of the $\SSymbols$-algebra $d\SAlgebra'_0$ with $S$-sorted set of carriers $\SemDomain$
together with interpretations $P^{d\SStructure'}_w$ (for $P\in\SPredicates_w$) defined to be 
\[\begin{array}{rcl}
P^{d\SStructure'}_w & = & \{([t_1]^\alpha_{d\SAlgebra'},\ldots,[t_n]^\alpha_{d\SAlgebra'})\mid (t_1,\ldots,t_n)\in\STerms_w, 
\alpha\in\Variables\to\SemDomain,\\
& & \hspace{1cm}d(P)=P'(t'_1,\ldots,t'_m), \cY=\Var(t'_1,\ldots,t'_m), \sigma(y_i)=t_i,  1\leq i\leq n\\
& & \hspace{1cm} \exists\alpha':\cY\to\SemDomain'([\sigma(t'_1)]^{\alpha'}_{\SAlgebra'},\ldots,[\sigma(t'_m)]^{\alpha'}_{\SAlgebra'})\in (P')^{\SStructure'}\}
\end{array}\]
\end{definition}
Note that $\langle\tau,d\rangle$ 
can be seen now as a transformation 
$d:\mathit{Form}_{\SSymbols,\SPredicates}\to\mathit{Form}_{\SSymbols',\SPredicates'}$:
\[\begin{array}{rcl}
d(P(t_1,\ldots,t_n)) & = & d(P)[y_1\mapsto d(t_1),\ldots,y_n\mapsto d(t_n)]\\
d(\neg\phi) & = & \neg d(\phi)\\
d(\phi\wedge\phi') & = & d(\phi)\wedge d(\phi')\\
d((\forall x:s)\phi) & = & (\forall x:\tau(s)) d(\phi)
\end{array}
\]
The following obvious result formalizes the use of the previous construction.

\begin{theorem}\label{TheoDerivedSatifaction}
Let $\SSymbols=(S,\leq,\SSymbols,\SPredicates)$ and $\SSymbols'={(S',\leq',\SSymbols',\SPredicates')}$ be order-sorted signatures with predicates and $\langle\tau,d\rangle$ be a derivor from ${(S,\leq,\SSymbols,\SPredicates)}$ to ${(S',\leq',\SSymbols',\SPredicates')}$.
Let $\SStructure'$ be an 
$(S',\leq',\SSymbols',\SPredicates')$-structure and $\varphi\in\mathsf{Form}_{\SSymbols,\SPredicates}$.
If $\SStructure'\models d(\varphi)$, then $d\SStructure'\models\varphi$.
\end{theorem}
The following corollary of Theorem \ref{TheoDerivedSatifaction} formalizes our approach of seeking models of theories 
through derived structures.

\begin{corollary}[Derived model]\label{CoroDerivedModel}
Let $\SSymbols=(S,\leq,\SSymbols,\SPredicates)$ and $\SSymbols'={(S',\leq',\SSymbols',\SPredicates')}$ be order-sorted signatures with predicates and $\langle\tau,d\rangle$ be a derivor from ${(S,\leq,\SSymbols,\SPredicates)}$ to ${(S',\leq',\SSymbols',\SPredicates')}$.
Let $\SStructure'$ be an 
$(S',\leq',\SSymbols',\SPredicates')$-structure and $\cS\subseteq\mathit{Forms}_{\SSymbols,\SPredicates}$ be a theory. 
If for all $\varphi\in\cS$, $\SStructure'\models d(\varphi)$, then $d\SStructure'\models\cS$.
\end{corollary}
The following example shows how to use Corollary \ref{CoroDerivedModel} 
together with an \emph{appropriate} derived model for proving termination of
the OS-TRS $\pr{ToyamaOS}$ in our running example.

\begin{example}\label{ExToyamaOS_ProofOfTermination}
For the OS-TRS in Figure \ref{FigOSFOLTheoryForToyamaOS}, we use a logical model with 
the derived algebra in Example \ref{ExToyamaOS_DerivedAlgebra}
and predicates $\to$ and $\to^*$ that are interpreted by $>$ and $\geq$ (over the naturals), respectively.
This model satisfies the sentences in Figure \ref{FigExToyamaOS_modelFormula} that translate 
the sentences (\ref{ExToyamaOS_sentence1})-(\ref{ExToyamaOS_sentence12}) in Figure \ref{FigOSFOLTheoryForToyamaOS}. 
\begin{figure}[t]
\begin{eqnarray}
\forall t\in\SemDomain_{\fpr{S}} & (t \geq t)\label{ExToyamaOS_interpFormula1}\\
\forall t\in\SemDomain_{\fpr{S1}} & (t \geq t)\label{ExToyamaOS_interpFormula2}\\
\forall t,t',u\in\SemDomain_{\fpr{S}} & (t>t'\wedge t'\geq s\Rightarrow t\geq s)\label{ExToyamaOS_interpFormula3}\\
\forall t,t',u\in\SemDomain_{\fpr{S1}} & (t>t'\wedge t'\geq s\Rightarrow t\geq s)\label{ExToyamaOS_interpFormula4}\\
\forall t_1,t'_1,t_2,t_3\in\SemDomain_{\fpr{S1}} & (t_1>t'_1 \Rightarrow t_1+t_2+t_3>t'_1+t_2+t_3)\label{ExToyamaOS_interpFormula5}\\
\forall t_1,t_2,t'_2,t_3\in\SemDomain_{\fpr{S1}} & (t_2>t'_2 \Rightarrow t_1+t_2+t_3>t_1+t'_2+t_3)\label{ExToyamaOS_interpFormula6}\\
\forall t_1,t_2,t_3,t'_3\in\SemDomain_{\fpr{S1}} & (t_3>t'_3 \Rightarrow t_1+t_2+t_3>t_1+t_2+t'_3)\label{ExToyamaOS_interpFormula7}\\
\forall t_1,t'_1,t_2\in\SemDomain_{\fpr{S1}} & (t_1>t'_1 \Rightarrow t_1+t_2+1>t'_1+t_2+1)\label{ExToyamaOS_interpFormula8}\\
\forall t_1,t_2,t'_2,t_3\in\SemDomain_{\fpr{S1}} & (t_2>t'_2 \Rightarrow t_1+t_2+1>t_1+t'_2+1) \label{ExToyamaOS_interpFormula9}\\
\forall x\in\SemDomain_{\fpr{S2}} & (0+1+x> x+x+x)\label{ExToyamaOS_interpFormula10}\\
\forall x,y\in\SemDomain_{\fpr{S1}} & (x+y+1> x)\label{ExToyamaOS_interpFormula11}\\
\forall x,y\in\SemDomain_{\fpr{S1}} & (x+y+1> y)\label{ExToyamaOS_interpFormula12}\\[-1cm]\nonumber
\end{eqnarray}
\caption{Derived sentences for the sentences in Figure \ref{FigOSFOLTheoryForToyamaOS}}
\label{FigExToyamaOS_modelFormula}
\end{figure}
The validity of 
(\ref{ExToyamaOS_interpFormula1})-(\ref{ExToyamaOS_interpFormula9}) and
(\ref{ExToyamaOS_interpFormula11})-(\ref{ExToyamaOS_interpFormula12}) 
is obvious because $\SemDomain_{\fpr{S}}=\SemDomain_{\fpr{S1}}=\naturals$
and by reflexivity and transitivity of $\geq$ and the fact that $>\subseteq\geq$.
With regard to (\ref{ExToyamaOS_interpFormula10}), it 
holds due to our specific choice for $\SemDomain_{\fpr{S2}}$:
since $\SemDomain_{\fpr{S2}}=\{0\}$, $x$ is restricted to take value $0$;
thus, the condition $0+1+x> x+x+x$ becomes $1>0$, which is trivially true.
Since $>$ is a well-founded relation over $\SemDomain_{\fpr{S}}$ and $\SemDomain_{\fpr{S1}}$, termination of
$\pr{ToyamaOS}$ is proved.
\end{example}
Note that the model in Example \ref{ExToyamaOS_ProofOfTermination} is based on a \emph{decidable theory}, 
namely, Presburger's arithmetic (see Example \ref{ExPresburgerArithmeticIsADecidableTheory}).
Also, note that the interpretation of the one-step rewriting predicate 
$\to$ has been \emph{chosen} to be a well-founded ordering, which is essential to conclude termination
of $\pr{ToyamaOS}$ from the fact that $\SModel$ is a model of the sentences in Figure \ref{FigExToyamaOS_modelFormula}.

\section{Constraint-solving and automation of the analyses}\label{SecConstraintSolvingApproachToAnalyses}

The \emph{automatic generation} of models for a theory $\cS$ is 
a 
\emph{bottom-up} process where things remain `unspecified' until an attempt to \emph{solve}
some constraints obtained from $\cS$ succeeds.
The \emph{solution} is then used to synthesize a structure which is (by construction) a model of $\cS$.
This is accomplished as follows:
\begin{enumerate}
\item\label{ItemParameterizationStep} 
The syntactic objects are given \emph{parametric interpretations} of  a given type, usually chosen according
to their amenability to automation.
For instance, function symbols
are given \emph{linear polynomials} $a_1x_1+a_2x_2+\cdots+a_kx_k+a_0$, where
$a_0,a_1,\ldots,a_k$ are \emph{parameters} which are assumed to be \emph{existentially quantified} in
any formula during the generation process and variables $x_1,\ldots,x_k$ (of sorts $s_1,\ldots,s_k$) range on 
the interpretation domains $\SemDomain_{s_i}$ for $1\leq i\leq k$.
\item\label{ItemConstraintDefinitionStep} \emph{Sentences} $\phi\in\cS$ are used to obtain a new
set $\cS^\sharp$ of \emph{parametric} sentences $\exists\phi^\sharp$ with existentially quantified parameters $a_1,\ldots,a_n$.
Such parameters range over appropriate (constraint solving) domains $D_1,\ldots,D_n$.

\item\label{ItemConstraintSolvingStep} Then, $\cS^\sharp$ is treated as a \emph{constraint} whose solutions 
$\sigma=\{a_i\mapsto d_i\mid 1\leq i\leq n\}$, with $d_i\in D_i$ for $1\leq i\leq n$,
make $\sigma(\phi^\sharp)$ (an instantiation of the parameters in $\phi^\sharp$) \emph{true}.
\end{enumerate}
In the realm of this paper, the \emph{parameterization step} (item (\ref{ItemParameterizationStep}) above)
is part of the definition of \emph{derivors} (Definitions \ref{DerivedAlgebra} and \ref{DerivedStructure}).

Then, as remarked in item (\ref{ItemConstraintDefinitionStep}) above, the original theory $\cS$ is transformed into
a \emph{derived theory} $\cS^\sharp$.
In this paper $\cS^\sharp$ consists of arithmetic sentences, using numeric orderings as predicates.
Actually, an important issue is handling parametric formulas containing \emph{implications} of the form
\begin{eqnarray}
\bigwedge_{j=1}^{p_i}e_{ij}\geq d_{ij}\Rightarrow e_i\geq d_i\label{ConditionalConstraintInAffineForm}
\end{eqnarray}
where for all $i\in\{1,\ldots,k\}$, $p_i>0$ and for all $j$, $1\leq j\leq p_i$, 
$e_{ij}$ and $e_i$ are \emph{linear  expressions}  of the form $\sum a_kx_k$ for numbers $a_k$ and variables $x_k$,
and $d_{ij},d_i\in\reals$. 
Implications following the format (\ref{ConditionalConstraintInAffineForm}) are said to be in \emph{affine form}.
They are obtained
as derived formulas from the theory at stake (e.g., the theory in Figure \ref{FigOSFOLTheoryForToyamaOS}).
In this setting, the Affine form of Farkas' Lemma considered in
\cite[Section 5.1]{LucMes_ModelsForLogicsAndConditionalConstraintsInAutomatedProofsOfTermination_AISC14}
is useful.
In general, given $\vec{c}\in\reals^n$ and $\beta\in\reals$, 
the affine form of Farkas' Lemma can be used to check whether a constraint $\vec{c}^T\vec{x}\geq\beta$ holds 
whenever $\vec{x}$ ranges on the set $S$ of solutions $\vec{x}\in\reals^n$ of a linear
system $A\vec{x}\geq\vec{b}$ of $k$ inequalities, i.e., $A$ is a matrix of $k$ rows and $n$ columns
and $\vec{b}\in\reals^k$. 
According to Farkas' Lemma, we have to \emph{find} a vector $\vec{\lambda}$ of
$k$ non-negative numbers $\vec{\lambda}\in\reals^k_0$ such that
$\vec{c}=A^T\vec{\lambda}$ and $\vec{\lambda}^T\vec{b}\geq\beta$.

Farkas' Lemma permits the \emph{removal} of all variables $\vec{x}$ and the transformation of the conditional constraint 
into a set of equalities and inequalities that, as indicated in item (\ref{ItemConstraintSolvingStep}) above, 
can be handled by means of tools for arithmetic \emph{constraint solving} like 
\multisolver\footnote{\url{http://zenon.dsic.upv.es/multisolver/}}. Then, we obtain a model for $\cS$.
The following section provides a complete account of this process using our running example.

\section{Order-sorted structures with convex domains}\label{SecOrderSortedStructuresWithConvexDomains}

The resolution of our running example (Example \ref{ExToyamaOS_ProofOfTermination}) 
shows that flexibility in the definition of domains $\SemDomain_s$ 
for sorts $s\in S$ 
is an asset: we have \emph{simultaneously used} (due to the presence of sorts) 
an infinite domain like $\naturals$ (which is typical in termination proofs)
and the finite domain  $\{0\}$.
In order to provide an appropriate computational basis to the \emph{automatic} definition of algebras and
structures that can be used in program analysis with order-sorted first-order specifications, we follow
\cite{LucMes_ModelsForLogicsAndConditionalConstraintsInAutomatedProofsOfTermination_AISC14}  
and focus on domains that are obtained as the solution of polynomial and specially 
\emph{linear} constraints.

\begin{definition}[Convex polytopic domain]{\rm \cite[Definition 1]{LucMes_ModelsForLogicsAndConditionalConstraintsInAutomatedProofsOfTermination_AISC14}}
\label{DefConvexPolytopicDomain}
Given a  matrix $\MatrixConvexDomain\in\reals^{m\times n}$, and $\VectorConvexDomain\in\reals^m$, the set 
$D(\MatrixConvexDomain,\VectorConvexDomain)=\{\vec{x}\in\reals^n\mid \MatrixConvexDomain\vec{x}\geq\VectorConvexDomain\}$
is called a \emph{convex polytopic domain}.
\end{definition}
In  Definition \ref{DefConvexPolytopicDomain},
vectors $\vec{x},\vec{y}\in\reals^n$ are \emph{compared}  using 
the \emph{coordinate-wise} extension of the ordering $\geq$ among \emph{numbers} (by abuse, we use the same
symbol): 
$\vec{x}=(x_1,\ldots,x_n)^T\geq(y_1,\ldots,y_n)^T=\vec{y} \text{ if and only if } x_1\geq y_1\wedge\cdots \wedge x_n\geq y_n$.
Convex domains 
can be parameterized by considering a subset $N\subseteq\reals$ (e.g., $\naturals$, $\integers$,
$\rationals$, etc.) with $\MatrixConvexDomain\in N^{m\times n}$, and $\VectorConvexDomain\in N^m$ and defining $D_N(\MatrixConvexDomain,\VectorConvexDomain)=\{\vec{x}\in N^n\mid \MatrixConvexDomain\vec{x}\geq\VectorConvexDomain\}$.
\begin{example}\label{ExIntendedInterpretationsOfSortsAsConvexDomains}
\emph{Intended} interpretations $\SemDomain^s$ for some usual sorts $s$ as convex domains $\SemDomain_s=D(\MatrixConvexDomain^s,\VectorConvexDomain^s)$ are:
\begin{center}
\begin{tabular}{|c|c|c|c|}
\hline
Sort & $\MatrixConvexDomain^s$ & $\VectorConvexDomain^s$ & $\SemDomain_s=D(\MatrixConvexDomain^s,\VectorConvexDomain^s)$\\
\hline\hline
$\emptyset$ & $(0)$ & $(1)$ & $\emptyset$\\
\hline
Nat & $(1)$ & $(0)$ & $[0,+\infty)$\\
\hline
NzNat & $(1)$ & $(1)$ & $[1,+\infty)$\\
\hline
Zero & $(1,-1)^T$ & $(0,0)^T$ & $\{0\}$\\
\hline
Bool & $(1,-1)^T$ & $(0,-1)^T$ & $[0,1]$\\
\hline
Char & $(1,-1)^T$ & $(0,-255)^T$ & $[0,255]$\\
\hline
\end{tabular}
\end{center}
\end{example}
We discuss the automatic generation of structures based on convex polytopic domains according to the 
general scheme in Section \ref{SecConstraintSolvingApproachToAnalyses}.
We illustrate the develoment by using our running example.

\subsection{Domains}\label{SecGeneratingDomains}

We interpret sorts $s\in S$  as convex domains 
$\SemDomain_s=D(\MatrixConvexDomain^s,\VectorConvexDomain^s)$, where\footnote{In the following, we use
write the sort $s$ in the superscript of the matrix and vector components $\MatrixConvexDomain$ and $\VectorConvexDomain$
of the convex domain. In this way, we can
use the subscripts to identify their \emph{components}: rows, columns, etc.}
$\MatrixConvexDomain^s\in\reals^{m_s\times n_s}$ is an $m_s\times n_s$-matrix
and $\VectorConvexDomain^s\in\reals^{m_s}$.
Thus, $\SemDomain_s\subseteq\reals^{n_s}$.
Given $s\in S$, we have to \emph{fix} $m_s$ and $n_s$ according to some
criterion.
Then, matrices $\MatrixConvexDomain^s$ and vectors $\VectorConvexDomain^s$ can be written 
\emph{parametrically}.
The exact shape of $D(\MatrixConvexDomain^s,\VectorConvexDomain^s)$ will be settled by the subsequent \emph{constraint solving process}.

\begin{remark}\label{RemDimensionOfConvexDomain}
For 1-dimensional convex domains $D(\MatrixConvexDomain^s,\VectorConvexDomain^s)\subseteq\reals$ (i.e., intervals, with $n_s=1$), imposing $0<m_s\leq 2$ is appropriate because the existence of more
than $2$ rows in $\MatrixConvexDomain^s$ for a given entry in $\VectorConvexDomain^s$ 
is useless: they define the same interval that
those producing the least and bigger values when applying them to $\vec{x}$.
In general, if $m_s=2$, then $\MatrixConvexDomain^s=(C^s_1,C^s_2)^T$ and $\VectorConvexDomain^s=(b^s_1,b^s_2)^T$ means that $C^s_1x\geq b^s_1$ 
and $C^s_2x\geq b^s_2$. 
As shown in Example \ref{ExIntendedInterpretationsOfSortsAsConvexDomains}, fixing $m_s=2$ and
using $\integers$ as domain for parameters $b_i$ and $c_i$ is important to gain flexibility in the definition
of convex domains, especially if bounded domains are desirable.
Our choice, in this 1-dimensional case is $m_s=2$ and $n_s=1$.
\end{remark}

\subsubsection{Non-empty convex domains}\label{SectionNonEmptyConvexDomains}

An important requirement in termination analysis is that the domain $D(\MatrixConvexDomain,\VectorConvexDomain)\subseteq\reals^n$ where a
well-founded relation $>$ is to be defined should \emph{not be empty}.
At the syntactic level we guarantee this by just adding a \emph{fresh} constant $\pr{k}$ of the appropriate sort 
$\pr{S}$ (to be interpreted by $D(\MatrixConvexDomain,\VectorConvexDomain)$) in the signature: $\pr{k:S}$.
Of course, if such a constant is already part of the specification, nothing else is required.
At the \emph{derived} level this becomes  a (vectorial) constraint $\MatrixConvexDomain k^T\geq\VectorConvexDomain$
to be satisfied by a \emph{dummy} constant $k\in\reals^n$.

\subsubsection{Convex domains which are bounded from below}\label{SecConvexDomainsBoundedFromBelow}

In some applications, it is useful to guarantee that a semantic domain $\SemDomain$ is \emph{bounded from below}.
In our setting, the following sentence (which is universally quantified on variable $x$):
\begin{eqnarray}
\MatrixConvexDomain x\geq\VectorConvexDomain\Rightarrow x\geq\vec{\alpha}\nonumber%
\end{eqnarray}
guarantees that $\SemDomain=D(\MatrixConvexDomain,\VectorConvexDomain)$ is  bounded from below;
here $\vec{\alpha}$ is a fresh \emph{constant} whose value will be established by the constraint solving process. 

\subsubsection{Compatibility with the subsort relation}\label{SecCompatibilityWithSubsortRelation}

Regarding the \emph{subsort} relation, if $s\leq s'$, then 
$\SemDomain_s=D(\MatrixConvexDomain^s,\VectorConvexDomain^s)\subseteq D(\MatrixConvexDomain^{s'},\VectorConvexDomain^{s'})=\SemDomain_{s'}$ 
must hold. 
Such a condition is expressed  by the universally quantified formula:
\begin{eqnarray}
\MatrixConvexDomain^sx\geq\VectorConvexDomain^s\Rightarrow \MatrixConvexDomain^{s'}x\geq\VectorConvexDomain^{s'}
\end{eqnarray}

\subsection{Functions}\label{SecConvexMatrixInterpretations}

A \emph{many-sorted convex matrix intepretation} for 
$f:s_1\cdots s_k\to s$ is a  linear expression $F_1x_1+\cdots+F_kx_k+F_0$ such that
(1) for all $i$, $1\leq i\leq k$, $F_i\in\reals^{n_s\times n_{s_i}}$
are $n_s\times n_{s_i}$-matrices and $x_i$ are variables ranging on $\reals^{n_{s_i}}$, 
(2) $F_0\in\reals^{n_s}$, 
and 
(3) it ranges on $D(\MatrixConvexDomain^s,\VectorConvexDomain^s)$ whenever variables $x_i$ take
value on the corresponding domain $D(\MatrixConvexDomain^{s_i},\VectorConvexDomain^{s_i})$, i.e., that satisfies 
the following \emph{algebraicity condition}:

{\small\begin{eqnarray}
\forall x_1\in\reals^{n_{s_1}},\ldots\forall x_k\in\reals^{n_{s_k}}\left (\bigwedge_{i=1}^k \MatrixConvexDomain^{s_i}x_i\geq \VectorConvexDomain^{s_i}\Rightarrow \MatrixConvexDomain^s(F_1x_1+\cdots+F_kx_k+F_0)\geq \VectorConvexDomain^s\right )\nonumber%
\end{eqnarray}}

For overloaded symbols 
$f\in\SSymbols_{w,s}\cap\SSymbols_{w',s'}$ with $w\leq w'$, we must have
$s\leq s'$ as well. We have to guarantee that the interpretations $f^\SAlgebra_{w,s}$ and $f^\SAlgebra_{w',s'}$ coincide
on $\SemDomain_{w}$ (see Section \ref{SecOrderSortedFirstOrderLogic})
As discussed in Section \ref{SecGeneratingDomains}, this implies that, with $w=s_1\cdots s_k$ and $w'=s'_1\cdots s'_k$,
we must have $n_{s_i}=n_{s'_i}$ for all $i$, $1\leq i\leq k$. Furthermore, $n_s=n_{s'}$ as well.
Therefore, if 
$f^\SAlgebra_{w,s}=\sum_{i=1}^k F_i\vec{x_i}+F_0$ and $f^\SAlgebra_{w',s'}=\sum_{i=1}^k F'_i\vec{x_i}+F'_0$,
the desired condition can be written as follows:
\begin{eqnarray}
\forall x_1\in\SemDomain_{s_1},\ldots,\forall x_k\in\SemDomain_{s_n}, \sum_{i=1}^k (F_i-F'_i)x_i+F_0-F'_0=0\hspace{0.5cm}\text{or, equivalently:}\nonumber\\
\forall x_1\in\reals^{n_{s_1}}\ldots,\forall x_k\in\reals^{n_{s_k}}(\bigwedge_{i=1}^k\MatrixConvexDomain^{s_i}x_i\geq\VectorConvexDomain^{s_i}\Rightarrow \sum_{i=1}^k (F_i-F'_i)x_i+F_0-F'_0=0)\nonumber
\end{eqnarray}

\subsection{Predicates}\label{SecGeneratingPredicates}

The interpretation of the (universally quantified) rules of the theory for the running example, 
with overloaded predicates $\to,\to^*$ (see Example \ref{ExToyamaOS_SignatureWithPredicates}), is given by interpreting the overloads of $\to^*$ as $\geq$ 
(the usual ordering on numbers) and the overloads of $\to$ as $>_\delta$ for some $\delta>0$.
The use of this special ordering over the reals instead of the usual one $>_\reals$ is due to the need of
interpreting $\to$ by using a \emph{well-founded ordering} in order to obtain a sound termination analysis.
According to \cite{Lucas_PolOverRealsTheoPrac_TIA05}, $>_\delta$ is well-founded over subsets $A\subseteq\reals$ that are \emph{bounded from below}. 

\section{Automatic treatment of the running example}\label{SecAutomaticTreatmentOfTheRunningExample}
Since we deal with three different sorts $\pr{S}$, $\pr{S1}$, and $\pr{S2}$, we consider three convex domains:
\[
\begin{array}{rcl@{\hspace{1cm}}rcl@{\hspace{1cm}}rcl}
\SemDomain_{\pr{S}} & = & D(\MatrixConvexDomain^S,\VectorConvexDomain^S) & \SemDomain_{\pr{S1}} & = & D(\MatrixConvexDomain^{S1},\VectorConvexDomain^{S1}) & \SemDomain_{\pr{S2}} & = & D(\MatrixConvexDomain^{S2},\VectorConvexDomain^{S2}) 
\end{array}
\]
where $\MatrixConvexDomain^S,\MatrixConvexDomain^{S1},\MatrixConvexDomain^{S2}\in\reals^{2\times 1}$ and 
$\VectorConvexDomain^S,\VectorConvexDomain^{S1},\VectorConvexDomain^{S2}\in\reals^2$.
By requiring non-emptyness, we obtain the first constraints for our running example:
\begin{eqnarray}
C^S_1 k\geq b^S_1\wedge C^S_2 k\geq b^S_2\label{ExToyamaOS_SortSNonEmpty}\\
C^{S1}_1 k'\geq b^{S1}_1\wedge C^{S1}_2 k'\geq b^{S1}_2\label{ExToyamaOS_SortS1NonEmpty}
\end{eqnarray}
where $k$ and $k'$ are \emph{dummy} elements $k,k'\in\reals$ for $\pr{S}$ and $\pr{S1}$.
However, since \verb$ToyamaOS$ already includes a symbol $\pr{1}$ of sort $\pr{S1}$,
constraint (\ref{ExToyamaOS_SortS1NonEmpty}) is not really necessary and could be avoided (see constraint (\ref{ExToyamaOS_ParametricAlgebraicInterpretation_AlgebraicityOfOne})
below).
And, although there is no constant symbol of sort $\pr{S}$, function $\pr{f}$ takes arguments of sort $\pr{S1}$ (which is not empty) 
and yields
a term of sort $\pr{S}$. Thus, sort $\pr{S}$ is not empty; this is guaranteed by means of other constraints like 
(\ref{ExToyamaOS_ParametricAlgebraicInterpretation_AlgebraicityOfFf1})-(\ref{ExToyamaOS_ParametricAlgebraicInterpretation_AlgebraicityOfFf2}) below.
Thus, (\ref{ExToyamaOS_SortSNonEmpty}) could be avoided too.

We guarantee that $\SemDomain_{\pr{S}}$ and $\SemDomain_{\pr{S1}}$ are both bounded from below with the following constraints:
\begin{eqnarray}
C^S_1x\geq b^S_1\wedge C^S_2 x\geq b^S_2 & \Rightarrow & x\geq\alpha\label{ExToyamaOS_ConditionForSboundedFromBelow}\\
C^{S1}_1 x\geq b^{S1}_1\wedge C^{S1}_2 x\geq b^{S1}_2 & \Rightarrow & x\geq\alpha'\label{ExToyamaOS_ConditionForS1boundedFromBelow}
\end{eqnarray}
for constants $\alpha$ and $\alpha'$, where $x$ is universally quantified (but $\alpha$ and $\alpha'$ are treated as new, existentially
quantified, parameters).
Since $\pr{S2}\leq\pr{S1}$,
we add the following sentence (universally quantified in $x$):
\begin{eqnarray}
C^{S2}_1x\geq b^{S2}_1\wedge C^{S2}_2x\geq b^{S2}_2\Rightarrow C^{S1}_1x\geq b^{S1}_1\wedge C^{S1}_2x\geq b^{S1}_2\nonumber 
\end{eqnarray}
However, since this sentence is \emph{not} in affine form (due to the conjunction in the consequent of the implication), we decompose it
as a conjunction of two implications as follows:
\begin{eqnarray}
C^{S2}_1x\geq b^{S2}_1\wedge C^{S2}_2x\geq b^{S2}_2\Rightarrow C^{S1}_1x\geq b^{S1}_1\label{SubsortCondition1}\\
C^{S2}_1x\geq b^{S2}_1\wedge C^{S2}_2x\geq b^{S2}_2\Rightarrow C^{S1}_2x\geq b^{S1}_2 \label{SubsortCondition2}
\end{eqnarray}

With regard to function symbols, since $n_s=1$ for all $s\in S$,
components $F_i$ for each symbol $f\in\SSymbols$ are \emph{numbers}, actually.
We give \emph{parametric interpretations} to each $f\in\SSymbols$ as follows:
\[
\begin{array}{rcl@{\hspace{0.7cm}}rcl@{\hspace{0.7cm}}rcl}
{}[\pr{0}] & = & z_0 & [\pr{1}] & = & u_0\\
{}[\pr{f}](x,y,z) & = & f_1x+f_2y+f_3z+f_0 & [\pr{g}](x,y) & = & g_1x+g_2y+g_0
\end{array}
\]
and the \emph{algebraicity conditions} are (with $x,y,z$ universally quantified in all formulas):
{\footnotesize
\begin{eqnarray}
C^{S2}_1z_0\geq b^{S2}_1\wedge C^{S2}_2z_0\geq b^{S2}_2\label{ExToyamaOS_ParametricAlgebraicInterpretation_AlgebraicityOfZero}\\
C^{S1}_1u_0\geq b^{S1}_1\wedge C^{S1}_2u_0\geq b^{S1}_2\label{ExToyamaOS_ParametricAlgebraicInterpretation_AlgebraicityOfOne}\\
\bigwedge_{i=1}^2C^{S1}_ix\geq b^{S1}_i \wedge
\bigwedge_{i=1}^2C^{S1}_iy\geq b^{S1}_i\wedge 
\bigwedge_{i=1}^2C^{S1}_iz\geq b^{S1}_i & \Rightarrow & C^S_1(f_1x+f_2y+f_3z+f_0)\geq b^{S}_1\label{ExToyamaOS_ParametricAlgebraicInterpretation_AlgebraicityOfFf1}\\
\bigwedge_{i=1}^2C^{S1}_ix\geq b^{S1}_i \wedge
\bigwedge_{i=1}^2C^{S1}_iy\geq b^{S1}_i\wedge 
\bigwedge_{i=1}^2C^{S1}_iz\geq b^{S1}_i & \Rightarrow & C^S_2(f_1x+f_2y+f_3z+f_0)\geq b^{S}_2~~~~~\label{ExToyamaOS_ParametricAlgebraicInterpretation_AlgebraicityOfFf2}\\
\bigwedge_{i=1}^2C^{S1}_ix\geq b^{S1}_i\wedge 
\bigwedge_{i=1}^2C^{S1}_iy\geq b^{S1}_i  & \Rightarrow & C^{S1}_1(g_1x+g_2y+g_0)\geq b^{S1}_1\label{ExToyamaOS_ParametricAlgebraicInterpretation_AlgebraicityOfFg1}\\
\bigwedge_{i=1}^2C^{S1}_ix\geq b^{S1}_i\wedge 
\bigwedge_{i=1}^2C^{S1}_iy\geq b^{S1}_i  & \Rightarrow & C^{S1}_2(g_1x+g_2y+g_0)\geq b^{S1}_2\label{ExToyamaOS_ParametricAlgebraicInterpretation_AlgebraicityOfFg2}
\end{eqnarray}
}%
where (\ref{ExToyamaOS_ParametricAlgebraicInterpretation_AlgebraicityOfFf1}) and (\ref{ExToyamaOS_ParametricAlgebraicInterpretation_AlgebraicityOfFf2}) are actually obtained from a single algebraicity condition after 
splitting the conjunction in the consequent of the
implication to obtain implications in affine form (as in Section \ref{SecCompatibilityWithSubsortRelation}). 
Similarly for 
(\ref{ExToyamaOS_ParametricAlgebraicInterpretation_AlgebraicityOfFg1}) and (\ref{ExToyamaOS_ParametricAlgebraicInterpretation_AlgebraicityOfFg2})).
Note also that, even though $\pr{0}$ and $\pr{1}$ are constant symbols, 
(\ref{ExToyamaOS_ParametricAlgebraicInterpretation_AlgebraicityOfZero}) and (\ref{ExToyamaOS_ParametricAlgebraicInterpretation_AlgebraicityOfOne}) are also necessary to guarantee that they receive
a value according to their \emph{sort} ($\pr{S2}$ and $\pr{S1}$, respectively).

Using the interpretations for sorts, function symbols, and predicates, we obtain the following derived sentences:
\begin{enumerate}
\item\label{ItemReflexivityRule} Instances of the \emph{reflexivity} rule (Rf), corresponding to sentences (\ref{ExToyamaOS_sentence1}) and 
(\ref{ExToyamaOS_sentence2}) in Figure \ref{FigOSFOLTheoryForToyamaOS}, 
with $t$ universally quantified:
\begin{eqnarray}
C^S_1t\geq b^S_1\wedge C^S_2t\geq b^S_2\Rightarrow t\geq t\label{ExToyamaOS_sentence1_Parametric}\\
C^{S1}_1t\geq b^{S1}_1\wedge C^{S1}_2t\geq b^{S1}_2\Rightarrow t\geq t\label{ExToyamaOS_sentence2_Parametric}
\end{eqnarray}
Note that the two sentences above trivially hold \emph{under the current interpretation of $\geq$ as a quasi-ordering} (a
\emph{reflexive} and \emph{transitive} relation). 
Thus, (\ref{ExToyamaOS_sentence1_Parametric}) and (\ref{ExToyamaOS_sentence2_Parametric}) could be \emph{removed}.
\item\label{ItemTransitivityRule} Instances of the \emph{transitivity} rule (T),
corresponding to (\ref{ExToyamaOS_sentence3}) and (\ref{ExToyamaOS_sentence4}): 
\begin{eqnarray}
\bigwedge_{i=1}^2C^S_it\geq b^S_i\wedge \bigwedge_{i=1}^2C^S_it'\geq b^S_i\wedge\bigwedge_{i=1}^2C^S_iu\geq b^S_i\wedge t\geq t'+\delta\wedge t'\geq u\Rightarrow t\geq u\label{ExToyamaOS_sentence3_Parametric}\\
\bigwedge_{i=1}^2C^{S1}_it\geq b^{S1}_i\wedge \bigwedge_{i=1}^2C^{S1}_it'\geq b^{S1}_i\wedge\bigwedge_{i=1}^2C^{S1}_iu\geq b^{S1}_i\wedge t\geq t'+\delta\wedge t'\geq u\Rightarrow t\geq u\label{ExToyamaOS_sentence4_Parametric}
\end{eqnarray}

\item Instances of the \emph{congruence} rule (C), 
corresponding to (\ref{ExToyamaOS_sentence5})-(\ref{ExToyamaOS_sentence9}), where we use $t\in\SemDomain_{\pr{S1}}$ instead of
$C^{S1}_1t\geq b^{S1}_1\wedge C^{S1}_2t\geq b^{S1}_2$: 
\begin{eqnarray}
\bigwedge_{i=1}^3t_i\in\SemDomain_{\pr{S1}}\wedge t'_1\in\SemDomain_{\pr{S1}}\wedge t_1\geq t'_1+\delta \Rightarrow f_1t_1+f_2t_2+f_3t_3+f_0\geq f_1t'_1+f_2t_2+f_3t_3+f_0+\delta\label{ExToyamaOS_sentence5_Parametric}\\
\bigwedge_{i=1}^3t_i\in\SemDomain_{\pr{S1}}\wedge t'_2\in\SemDomain_{\pr{S1}}\wedge t_2\geq t'_2+\delta \Rightarrow f_1t_1+f_2t_2+f_3t_3+f_0\geq f_1t_1+f_2t'_2+f_3t_3+f_0+\delta\label{ExToyamaOS_sentence6_Parametric}\\
\bigwedge_{i=1}^3t_i\in\SemDomain_{\pr{S1}}\wedge t'_3\in\SemDomain_{\pr{S1}}\wedge t_3\geq t'_3+\delta \Rightarrow f_1t_1+f_2t_2+f_3t_3+f_0\geq f_1t_1+f_2t_2+f_3t'_3+f_0+\delta\label{ExToyamaOS_sentence7_Parametric}\\
\bigwedge_{i=1}^2t_i\in\SemDomain_{\pr{S1}}\wedge t'_1\in\SemDomain_{\pr{S1}}\wedge t_1\geq t'_1+\delta \Rightarrow g_1t_1+g_2t_2+g_0\geq g_1t'_1+g_2t_2+g_0+\delta\label{ExToyamaOS_sentence8_Parametric}\\
\bigwedge_{i=1}^2t_i\in\SemDomain_{\pr{S1}}\wedge t'_2\in\SemDomain_{\pr{S1}}\wedge t_2\geq t'_2+\delta \Rightarrow g_1t_1+g_2t_2+g_0\geq g_1t_1+g_2t'_2+g_0+\delta\label{ExToyamaOS_sentence9_Parametric}
\end{eqnarray}

\item Instances of the \emph{replacement} rule (Re), corresponding to (\ref{ExToyamaOS_sentence10})-(\ref{ExToyamaOS_sentence11}):
\begin{eqnarray}
C^{S2}_1 x\geq b^{S2}_1\wedge C^{S2}_2 x\geq b^{S2}_2\Rightarrow f_1z_0+f_2u_0+f_3x+f_0\geq f_1x+f_2x+f_3x+f_0+\delta\label{ExToyamaOS_sentence10_Parametric}\\
C^{S1}_1 x\geq b^{S1}_1\wedge C^{S1}_2 x\geq b^{S1}_2\wedge C^{S1}_1 y\geq b^{S1}_1\wedge C^{S1}_2 y\geq b^{S1}_2\Rightarrow g_1x+g_2y+g_0\geq x+\delta\label{ExToyamaOS_sentence11_Parametric}\\
C^{S1}_1 x\geq b^{S1}_1\wedge C^{S1}_2 x\geq b^{S1}_2\wedge C^{S1}_1 y\geq b^{S1}_1\wedge C^{S1}_2 y\geq b^{S1}_2\Rightarrow g_1x+g_2y+g_0\geq y+\delta\label{ExToyamaOS_sentence12_Parametric}
\end{eqnarray}
\end{enumerate}

\subsection{Synthesis of the model}

The conjunction of all previous sentences (\ref{ExToyamaOS_SortSNonEmpty})-(\ref{ExToyamaOS_sentence12_Parametric}) 
(perhaps dropping some of them, as suggested in previous sections)
yields an $\exists\forall$-sentence (the $\exists$ concerns existential quantification of $k$, $k'$, $\alpha$, $\alpha'$, $\delta$, and all parameters
in domain descriptions and algebraic interpretations) where all \emph{introduced parameters} 
are existentially quantified (on appropriate domains of coefficients, see Section \ref{SecConstraintSolvingApproachToAnalyses})
and all \emph{semantic variables} (i.e., those ultimately coming from the description of the problem and required by the semantic
interpretation of symbols) 
are universally quantified (over the reals).
As mentioned in Section \ref{SecConstraintSolvingApproachToAnalyses}, we can use now the techniques discussed in \cite{LucMes_ModelsForLogicsAndConditionalConstraintsInAutomatedProofsOfTermination_AISC14}
together with standard constraint solving techniques to obtain an
assignment of values to the parameters which defines the desired model. 
Given  a matrix $A$ of $k$ rows and $n$ columns,
$\vec{b}\in\reals^k$,
$\vec{c}\in\reals^n$ and $\beta\in\reals$,
the application of the affine form of Farkas' Lemma to prove that the universally quantified sentence 
$A\vec{x}\geq\vec{b}\Rightarrow\vec{c}^T\vec{x}\geq\beta$ holds
tries to \emph{find} a vector $\vec{\lambda}$ of
$k$ non-negative numbers $\vec{\lambda}\in\reals^k_0$ such that the \emph{constraints}
$\vec{c}=A^T\vec{\lambda}$ and $\vec{\lambda}^T\vec{b}\geq\beta$ hold.

\begin{example}
We apply the Affine form of Farkas' Lemma to sentence (\ref{ExToyamaOS_ConditionForSboundedFromBelow})
as follows: the associated matrix $A$ is actually 
a vector $(C^S_1,C^S_2)^T$ and $\vec{b}=(b^S_1,b^S_2)^T$; we have that $\vec{c}=(1)^T$ is a one-dimensional vector 
and finally $\beta=\alpha$. 
Then,  we seek a vector $\vec{\lambda}=(\lambda_1,\lambda_2)^T$ with $\lambda_1,\lambda_2\geq 0$ that satisfies
the (in)equations:
\[\begin{array}{r@{\:}c@{\:}l@{\hspace{1cm}}r@{\:}c@{\:}l@{\hspace{1cm}}l}
1 & = & C^S_1\lambda_1+C^S_2\lambda_2 & \lambda_1b^S_1+\lambda_2b^S_2 & \geq & \alpha & \lambda_1,\lambda_2\geq 0
\end{array}
\]
The satisfiability of these inequations (a constraint solving problem for parameters $C^S_1$, $C^S_2$, $b^S_1$, $b^S_2$,
$\lambda_1$, $\lambda_2$ and $\alpha$), is equivalent to the satisfiability of (\ref{ExToyamaOS_ConditionForSboundedFromBelow}).
\end{example}

\begin{example}
Sentence (\ref{ExToyamaOS_sentence12_Parametric}) is \emph{not} in \emph{affine} form, but we can easily fix it
as follows: 
\begin{eqnarray}%
C^{S1}_1 x\geq b^{S1}_1\wedge C^{S1}_2 x\geq b^{S1}_2\wedge C^{S1}_1 y\geq b^{S1}_1\wedge C^{S1}_2 y\geq b^{S1}_2\Rightarrow g_1x+(g_2-1)y\geq \delta-g_0\label{ExToyamaOS_sentence12_ParametricAffineForm}
\end{eqnarray}
Now, we apply Farkas' lemma to each of them.
The associated matrix $A$ has four rows (corresponding to the four atoms in the conjunction of the antecedent of the implication)
and two columns (corresponding to variables $x$ and $y$): 
$A=(C^{S1}_1,0~;~C^{S1}_2,0~;~0, C^{S1}_1~;~0, C^{S1}_2)$.
Vector $\vec{b}$ has four components: $\vec{b}=(b^{S1}_1,b^{S1}_2,b^{S1}_1,b^{S1}_2)^T$.
Now, $\vec{c}=(g_1,g_2-1)^T$ and $\beta=\delta-g_0$.
Thus, we want now a vector $\vec{\lambda}=(\lambda_1,\lambda_2,\lambda_3,\lambda_4)^T$  that satisfies:
\[\begin{array}{r@{\:}c@{\:}l@{\hspace{1cm}}r@{\:}c@{\:}l@{\hspace{1cm}}r@{\:}c@{\:}l@{\hspace{1cm}}r@{\:}c@{\:}l}
g_1 & = & C^{S1}_1\lambda_1+C^{S1}_2\lambda_2 & g_2-1 & = & C^{S1}_1\lambda_3+C^{S1}_2\lambda_4 \\ 
\lambda_1b^{S1}_1+ \lambda_2b^{S1}_2+ \lambda_3b^{S1}_1+ \lambda_4b^{S1}_2 & \geq & \delta-g_0 & \lambda_1,\lambda_2,\lambda_3,\lambda_4 & \geq & 0

\end{array}
\]
for some values of the parameters.
\end{example}
\begin{remark}
Note that each implication processed using Farkas' Lemma 
can use a \emph{different} vector $\vec{\lambda}$, 
but we have to \emph{solve a single set of inequations corresponding to a single solution which produces a single
model that makes all sentences valid}.
\end{remark}
The following assignment:
\[
\begin{array}{r@{\:}c@{\:}l@{\hspace{0.3cm}}r@{\:}c@{\:}l@{\hspace{0.3cm}}r@{\:}c@{\:}l@{~~}r@{\:}c@{\:}l@{\hspace{0.3cm}}r@{\:}c@{\:}l@{\hspace{0.3cm}}r@{\:}c@{\:}l@{\hspace{0.3cm}}r@{\:}c@{\:}l@{\hspace{0.3cm}}r@{\:}c@{\:}l@{\hspace{0.3cm}}r@{\:}c@{\:}l@{\hspace{0.3cm}}r@{\:}c@{\:}l@{\hspace{0.3cm}}r@{\:}c@{\:}l@{\hspace{0.3cm}}r@{\:}c@{\:}l@{\hspace{0.3cm}}r@{\:}c@{\:}l@{\hspace{0.3cm}}r@{\:}c@{\:}l@{~~}}
C^{S}_1 & = & 1 & C^S_2 & = & 1 & C^{S1}_1 & = & 1 & C^{S1}_2 & = & 1 & C^{S2}_1 & = & 1 & C^{S2}_2 & = & -1\\[0.2cm]
b^{S}_1 & = & 0  & b^{S}_2 & = & 0 & b^{S1}_1 & = & 0  & b^{S1}_2 & = & 0 & b^{S2}_1 & = & 0  & b^{S2}_2 & = & 0\\[0.2cm]
f_1 & = & 1 & f_2 & = & 1 & f_3 & = & 1 &  f_0 & = & 0 & g_1 & = & 1 & g_2 & = & 1 & g_0 & = & 0 & z_0 & = & 0 & u_0 & = & 1\\[0.2cm]
k & = & 0 & k' & = & 0 & \alpha & = & 0 & \alpha' & = & 0 & \delta & = & 1
\end{array}
\]
(where we disregard the different $\vec{\lambda}$ required by the application of Farkas' Lemma as \emph{administrative}
symbols) makes all sentences \emph{true} and \emph{generates} the model $\SModel$ for the theory $\cS$ in our
running example:
\[
\begin{array}{@{\!\!}r@{\:}c@{\:}l@{\hspace{0.7cm}}r@{\:}c@{\:}l@{\hspace{0.3cm}}r@{\:}c@{\:}l@{\hspace{0.5cm}}r@{\:}c@{\:}l@{\hspace{0.3cm}}r@{\:}c@{\:}l@{~~}r@{\:}c@{\:}l}
\SemDomain_{\pr{S}} & = & [0,+\infty) & \SemDomain_{\pr{S1}} & = & [0,+\infty) & \SemDomain_{\pr{S2}} & = & \{0\}\\[0.2cm]
\pr{f}^\SModel_{\pr{S1}\:\pr{S1}\:\pr{S1},\pr{S}}(x,y,z) & = & x+y+z & \pr{g}^\SModel_{\pr{S1}\:\pr{S1},\pr{S1}}(x,y) & = & x+y+1 & \pr{0}^\SModel_{\lambda,\pr{S2}} & = & 0 & \pr{1}^\SModel_{\lambda,\pr{S1}} & = & 1\\[0.2cm]
t\:\to^\SModel_{\pr{S}\:\pr{S}}\:t' & \Leftrightarrow & t>_1 t' &   t\:(\to^*)^\SModel_{\pr{S}\:\pr{S}}\: t' & \Leftrightarrow & t\geq t' & t\:\to^\SModel_{\pr{S1}\:\pr{S1}}\: t' & \Leftrightarrow & t>_1 t' & t\:(\to^*)^\SModel_{\pr{S1}\:\pr{S1}}\: t' & \Leftrightarrow & t\geq t'
\end{array}
\]

\section{Related work and conclusions}\label{sec:concl}

Our extension of derived algebras \cite{GogThaWag_AnInitialAlgebraApproachToSpecificationCorrectnessAndImplementationOfADTs_1978} to
derived models for order-sorted first-order theories follows some of the ideas in \cite{GogMes_ModelsAndEqualityForLogicalProgramming_TAPSOFT87}.
The generation of \emph{homogeneous algebras} 
using parametric interpretations followed by a constraint solving process
is standard in termination analysis of term rewriting \cite{ConMarTomUrb_MechProvTermPolInterp_JAR06}.
However, no systematic treatment of the generation of \emph{domains for sorts} and \emph{heterogeneous}
functions for \emph{ranked symbols}  in many-sorted or order-sorted algebras has been attempted to date.
And the generation of predicates as part of the generation of a model is also new.
This work is also a step forward in the practical use of logical models in proofs of operational termination of programs.
This was a main motivation of \cite{LucMes_ModelsForLogicsAndConditionalConstraintsInAutomatedProofsOfTermination_AISC14}
after understanding the practical role of using models in proofs of termination in the OT-Framework 
\cite{LucMes_ProvingOperationalTerminationOfDeclarativeProgramsInGeneralLogics_PPDP14,%
LucMes_LocalOperationalTerminationInGeneralLogics_FWirsing15}.
This paper also generalizes our previous experience in termination to envisage a generic, 
logic-oriented approach to abstraction in program analysis, which is based on 
defining appropriate \emph{models} for the logic which is used to describe the computations.
Focusing on an order-sorted first-order logic to describe programs and program properties, we 
have generalized the \emph{convex domains} and \emph{convex matrix interpretations} introduced in \cite{LucMes_ModelsForLogicsAndConditionalConstraintsInAutomatedProofsOfTermination_AISC14}
to the order-sorted setting.
Such a generalization leads to a flexible framework to define different domains for different sorts whereas it is
still amenable for automation by using existing algorithms and techniques from linear algebra \cite{Schrijver_TheoryOfLinearAndIntProg_1986}.
Indeed, the use of \emph{bounded} convex domains for some sorts (as $\{0\}$ for sort 
$\pr{S2}$ in $\pr{ToyamaOS}$) has been essential to
obtain a simple  solution of the corresponding problem.
A first implementation of the techniques presented in this paper has been reported in
\cite{Reinoso_LogicalModelsForAutomatedSemanticsDirectedProgramAnalysis_TFM15}, including
the generation of convex domains and convex interpretations along the lines of Section \ref{SecOrderSortedStructuresWithConvexDomains}.
The use of convex domains in termination analysis is also available as part of
the tool \muterm\ \cite{AlaGutLucNav_ProvingTerminationPropertiesWithMUTERM_AMAST10}.
Their usefulness has been recently shown in the 2015 International Termination Competition
held in August as part of CADE 2015, where convex domains have been successfully used to prove 
operational termination of \emph{conditional term rewriting systems}.

\bibliographystyle{eptcs}
\bibliography{biblio}
\end{document}